\documentclass[epj]{svjour}
\usepackage{ulem}

\usepackage{latexsym}
\usepackage{amssymb}
\usepackage{amsfonts}
\usepackage{amsmath}
\usepackage{bm}
\usepackage{graphicx}   
\usepackage{color}
\usepackage{subfigure}
\usepackage{times}
\usepackage{units}
\usepackage{lipsum}
\usepackage{hyperref}
\usepackage{multirow}
\usepackage{comment}

\graphicspath{ {./figures/} }

\newcommand{\Eref}[1]{Eq.~\eqref{#1}} 
\newcommand{\Fref}[1]{Fig.~\ref{#1}} 
\newcommand{\Sref}[1]{Sec.~\ref{#1}} 
\newcommand{\Tref}[1]{Table~\ref{#1}} 

\usepackage{color}
\definecolor{cyan}{rgb}{0,0.9,0.9}
\definecolor{orange}{rgb}{0.9,0.5,0}
\definecolor{magenta}{rgb}{1,0,1}
\definecolor{purple}{rgb}{0.8,0.4,0.8}
\definecolor{gray}{rgb}{0.8242,0.8242,0.8242}

\begin{document}
\title{Thermodynamics conditions of matter in the neutrino decoupling region during neutron star mergers}
\author{Andrea Endrizzi\inst{1} \and 
        Albino Perego\inst{2,3} \and
        Francesco M. Fabbri\inst{1} \and 
        Lorenzo Branca\inst{4} \and 
        David Radice \inst{5,6} \and 
        Sebastiano Bernuzzi \inst{1} \and 
        Bruno Giacomazzo \inst{4,7} \and
        Francesco Pederiva \inst{2,7} \and 
        Alessandro Lovato \inst{7,8}
        } 
%
\offprints{andrea.endrizzi@uni-jena.de}          
\institute{Theoretisch-Physikalisches Institut, Friedrich-Schiller-Universit\"at Jena, 07743, Jena, Germany \and
           Dipartimento di Fisica, Universit\`{a} di Trento, Via Sommarive 14, 38123 Trento, Italy \and 
           Istituto Nazionale di Fisica Nucleare, Sezione di Milano-Bicocca, Piazza della Scienza 20100, Milano, Italy \and 
           Dipartimento di Fisica G. Occhialini, Universit\`a di Milano - Bicocca, Piazza della Scienza 3, I-20126 Milano, Italy \and           
           Institute for Advanced Study, 1 Einstein Drive, Princeton, NJ 08540, USA \and 
           Department of Astrophysical Sciences, Princeton University, 4 Ivy Lane, Princeton, NJ 08544, USA \and
           INFN-TIFPA, Trento Institute for Fundamental Physics and Applications, Via Sommarive 14, I-38123 Trento, Italy \and
           Physics Division, Argonne National Laboratory, Argonne, IL 60439, USA}
\date{Received: 18 August 2019 / Revised version: 11 November 2019}
%
\abstract{
In this work we investigate the thermodynamics conditions at which neutrinos decouple from matter in neutron star merger remnants by post-processing results of merger simulations. We find that the matter density and the neutrino energies are the most relevant quantities in determining the decoupling surface location. For mean energy neutrinos 
($\sim$ 9, 15 and 24 MeV for $\nu_e$, $\bar{\nu}_e$ and $\nu_{\mu,\tau}$, respectively) the transition between diffusion and free-streaming conditions occurs around $10^{11}{\rm g~cm^{-3}}$ for all neutrino species. Weak and thermal equilibrium freeze-out occurs deeper (several $10^{12}{\rm g~cm^{-3}}$) for heavy flavor neutrinos than for $\bar{\nu}_e$ and $\nu_e$ ($\gtrsim 10^{11}{\rm g~cm^{-3}}$). Decoupling temperatures are broadly in agreement with the average neutrino energies, with softer equations of state characterized by $\sim$1 MeV larger decoupling temperatures. Neutrinos streaming at infinity with different energies come from different remnant parts. While low energy neutrinos ($ \sim 3~{\rm MeV}$) decouple at $ \rho \sim 10^{13}{\rm g~cm^{-3}}$, $T \sim 10~{\rm MeV}$ and $Y_e \lesssim 0.1$ close to weak equilibrium, high energy ones ($ \sim 50~{\rm MeV}$) decouple from the disk at $\rho \sim 10^{9}{\rm g~cm^{-3}}$, $T \sim 2~{\rm MeV}$ and $Y_e \gtrsim 0.25$. The presence of a massive NS or a BH influences the neutrino thermalization. While in the former case decoupling surfaces are present
for all relevant energies, the lower maximum density ($\lesssim 10^{12}{\rm g~cm^{-3}}$) in BH-torii systems does not allow softer neutrinos to thermalize and diffuse.
} 
\PACS{
  {97.60.Jd}{Neutron stars} \and
  {04.30.Db}{gravitational wave generation and sources} \and
  {04.25.D}{numerical relativity} \and  
  {97.60.Lf}{black holes (astrophysics)}
} 
\maketitle

	
\section{Introduction}
\label{sec:intro}

The merger of two neutron stars (NSs) results in the 
emission of a burst of gravitational waves (GWs), in the formation of
a massive remnant (possibly collapsing to a black hole, BH) and in the
ejection of matter in the interstellar medium, see \cite{Rosswog:2015nja,Baiotti:2016qnr,Barack:2018yly} for some recent reviews. 
The remnant is usually formed by a central object surrounded by a massive disk. The properties of the disk, as well as the timescale for BH collapse of the central object (if gravitationally unstable), depend mainly on the NS masses and on the still uncertain nuclear equation of state (EOS), see {\it e.g.}\cite{Hotokezaka:2011dh,Bauswein:2013jpa,Hotokezaka:2013iia,Zappa:2017xba,Radice:2018pdn,Dietrich:2018phi,Koppel:2019pys}.
Due to its high neutron content, the ejecta expelled in the interstellar medium can immediately synthetize heavy elements via the so-called rapid neutron capture process ($r$-process) nucleosynthesis \cite{Lattimer:1974a,Symbalisty:1982a,Thielemann:2017acv}.
The radioactive decay of these freshly synthetized elements powers a peculiar UV/optical/NIR electromagnetic transient called kilonova on a timescale ranging from a few hours up to months after the merger \cite{Li:1998bw,Kulkarni:2005jw,Metzger:2016pju}. Moreover, the merger remnant has been long thought to be the central engine of short-hard gamma-ray bursts, one of the most energetic and elusive extragalactic events detected up to cosmological distances \cite{Paczynski:1986px,Eichler:1989ve}.

This picture was recently confirmed by the detection of GW170817, the first GW signal from a binary neutron star (BNS) merger reported by LIGO and Virgo collaborations \cite{TheLIGOScientific:2017qsa}. 
Indeed, this detection was followed by the observation of
a short gamma-ray burst, GRB170817A (although a very weak one, due to
the off-axis relative orientation of the jet with respect to the
Earth, \cite{Mooley:2018dlz,Ghirlanda.etal:2019}) and of a kilonova (AT2017gfo) produced by the merger \cite{Monitor:2017mdv,GBM:2017lvd}.
During the first day, the kilonova was characterized by an intense quasi-thermal emission peaking in the UV and visible part of the electromagnetic spectrum, while on longer timescales it became redder and peaked at NIR frequences. Modelling of the associated light curves revealed the presence of more than one component in the ejecta, see {\it e.g.} \cite{Kasen:2017sxr,Villar:2017wcc,Perego:2017wtu,Wollaeger:2017ahm,Kawaguchi:2018ptg,Miller:2019dpt}. The components differ by their physical properties and possibly by their non-isotropic angular distributions. 
The low photon opacity required to explain the early kilonova emission suggests the presence of matter that was significantly irradiated by neutrinos. In this case, if the electron fraction increases above $\sim 0.25$ for a significant fraction of the ejecta then the production of $r$-process elements between the second and the third $r$-process peak is strongly suppressed. This observation has thus revealed the potential impact of neutrinos and weak interactions on the outcome and on the observables associated with BNS mergers.

During the coalescence the production of neutrinos of all flavors is boosted by the large temperatures (up to one hundred MeV, \cite{Perego:2019adq}) reached by dense matter, once a significant fraction of the kinetic energy of the two NSs has been converted into internal energy, see {\it e.g.} \cite{Eichler:1989ve,Ruffert:1996by,Rosswog:2003rv,Perego:2017fho}. Deep inside the forming hot remnant, for densities in excess of $10^{12}{\rm g~cm^{-3}}$, the neutrino mean free path for absorption and scattering processes becomes smaller than any relevant lenghtscale over which termodynamics quantities change significantly. Thus neutrinos form a trapped gas, in equilibrium with the plasma, that diffuses out on the diffusion timescale. When absorption reactions and inelastic scattering processes become less effective, weak and thermal equilibrium freezes out.
Neutrinos can still diffuse, due to the relatively large contribution of quasi-elastic scattering off free baryons to the total opacity. Only when lower densities ($\sim 10^{11}{\rm g~cm^{-3}}$) have been reached,  radiation can finally stream out freely. The region where the decoupling occurs is often referred as neutrino surface, or neutrinosphere in analogy with the photosphere of approximately spherically symmetric systems emitting photons, {\it e.g.} stars. Even in free-streaming conditions, neutrinos can still irradiate matter in the low density part of the remnant and in the expanding ejecta, potentially altering its neutron-to-proton content through charged-current reactions involving electron neutrinos and antineutrinos,  see {\it e.g.} \cite{Dessart:2008zd,Perego:2014fma,Just:2014fka,Foucart:2015gaa,Sekiguchi:2016bjd,Martin:2017dhc}.

Neutrino cross sections depend significantly on the incoming neutrino
energy: for energies in the range 1-100~MeV the dependence is
approximately quadratic \cite{Dicus:1972yr,Bruenn:1985en,Formaggio:2013kya}. The relevance of this dependence is twofold. On the one hand, neutrinos of different energies decouple at different locations inside the remnant. Thus neutrino spectra recorded far from the merger site result from matter-neutrino interaction over a wide range of thermodynamics conditions. On the other hand, the properties of the expanding ejecta, and ultimately of the resulting kilonova, depends significantly on the electron (anti)neutrino spectra emerging from the remnant through the effect of the free streaming neutrino absorption on the ejecta composition, see {\it e.g. }~\cite{Martin:2015hxa,Metzger:2014ila,Metzger:2017wot}.

The multidimensional nature of compact binary mergers
has represented a challenge in the production of quantitative
models. Only in the last three decades multidimensional numerical
simulations have shed light on the merger dynamics and on its many
observables. The inclusion of weak interactions in the context of BNS
mergers and of their aftermath has been done at different levels of
sophistications, including light bulbs, leakage schemes, moment
schemes, and Monte Carlo approaches, see {\it e.g.} \cite{Galeazzi:2013mia,Palenzuela:2015dqa,Foucart:2015gaa,Foucart:2016rxm,Perego:2015agy,Radice:2016dwd,Bovard:2017mvn,Perego:2017xth,Foucart:2018gis,Radice:2018pdn,Ardevol-Pulpillo:2018btx}. In most of the cases, grey ({\it i.e.} energy integrated) treatments have been adopted, while spectral treatments are still less common due to their significantly higher computational costs.
One of the key ingredients of any treatment of weak interactions are the neutrino opacities and their relation with the thermodynamical properties provided by the nuclear EOS.
While in the context of core-collapse supernovae (CCSNe) and proto-neutron stars (PNSs) many sensitivities studies on the relevant neutrino physics have been performed, {\it e.g.} \cite{Mezzacappa:1993gn,Roberts.etal:2012,Lentz:2012xc,Abdikamalov:2014oba,Melson:2015,Fischer:2016,OConnor:2018sti,Pan:2018vkx,Cabezon:2018lpr,Just.etal:2018}, very little have been said in the case of compact binary mergers. In the latter scenario the different physical conditions, as well as the different geometry of the system, can provide qualitatively different answers compared to the former. 
First studies about the properties of the neutrino surfaces in BNS
merger remnants were presented in \cite{Ruffert:1996by,Rosswog:2003rv}. In these works, the location of grey neutrino surfaces was determined by evaluating the optical depth along a finite number of directions, usually assuming cylindrical symmetry. A similar approach was adopted also by \cite{Perego:2014fma}, but in this case both the locations of dynamical and thermal decoupling were computed for different neutrino energies.
More sophisticated algorithms to compute neutrino optical depths
in multidimensional context have been developed, {\it e.g.} \cite{Perego:2014qda}, but never applied to detailed studies.
We recall that the evaluation of the optical depth in radiation hydrodynamical simulations is also required in a large number of approximated neutrino treatments \cite{Galeazzi:2013mia,Foucart:2015vpa,Perego:2015agy,Foucart:2016rxm}.

In this work, we present a first quantitative study of the the thermodynamics conditions of matter at the neutrinosphere in BNS remnants. Neutrinos opacities and optical depths are calculated by postprocessing numerical relativity simulations employing different nuclear EOSs and producing either NS or BH remnants. We identify the regions where the decoupling occurs and compute there the properties of the matter. 
In our analysis, we distinguish between the neutrino surfaces that denote the transition from the diffusion to the free-streaming regime, and the ones where weak and thermal equilibrium freezes out.

%
%

The paper is structured as follows: 
In \Sref{sec:scheme} we introduce the concepts of optical depth and the neutrino opacities used in this work. Then we present our algorithm for the optical depth evaluation and the analysis procedures followed in this work.
In \Sref{sec:results} we present the results obtained in terms of the distribution of thermodynamic quantities in the neutrino decoupling region, both for mean energy neutrinos and for several neutrino energies.
Finally, in \Sref{sec:conclusions} we summarize our results.

	

\section{Method}
\label{sec:scheme}

\subsection{The optical depth}

To evaluate the region where neutrinos decouple from matter, it is necessary to compute the optical depth for neutrinos at any point inside the simulation domain.
The optical depth quantifies the global degree of opacity of matter to 
radiation between two points ($A$ and $B$) along a certain path. If $\gamma$ denotes
the path between $A$ and $B$, the optical depth is defined as:
\begin{equation}
\label{eq:general optical depth}
\tau_{\gamma: A \rightarrow B} 
= \int_{\gamma: A \rightarrow B}
\kappa(s)~{\rm d} s
\end{equation} 
\noindent where $\kappa(s)$ is the opacity (corresponding to the inverse of the local mean free path) and ${\rm d}s = \sqrt{g_{ij} {\rm d}x^i {\rm d}x^j}$ is an infinitesimal displacement along the chosen path, with $g_{ij}$ being the 
local spatial metric.
The physical interpretation of $\tau$ emerges from its definition: $\tau$ counts the average
number of interactions that radiation particles experiences along $\gamma$.
In radiation transport problems involving astrophysical objects, radiation is often produced
in opaque regions from which it diffuses out on the diffusion timescale towards optically
thin regions located at the boundary of the physical system. In this
context, a statistical description of the radiation and of its global
behaviour is more relevant than the behaviour of single radiation
particles.
Even though radiation is often produced isotropically on the microscopic scales, 
interaction with matter can change the propagation direction on macroscopic scales. 
On the one hand, if radiation particles move towards a region of increasing mean free path, 
they will more probably retain their direction and eventually move away freely from the production site. On the other hand, particles moving towards a region of decreasing mean 
free path will more likely interact with matter, changing their original propagation 
direction. This implies that, independently from the emission properties, 
macroscopically radiation moves preferentially towards regions of larger mean free path.
Eventually, radiation particles reach regions where the local mean free 
path becomes larger than the relevant domain size and they will stream out of the system, practically without further interactions.
If we are interested in the global radiation behavior, we consider any point inside 
the domain as starting point ($A$) while the final point ($B$) can be any point on 
the boundary of the physical domain.

According to the statistical interpretation presented above, among all the possible 
paths connecting $A$ to the boundary, the most likely ways for radiation to escape
are the paths that minimize the optical depth.
We thus define the optical depth of a point $\mathbf{x}$ as
\begin{equation}
\label{eq:tau x}
\tau(\mathbf{x})=\min_{\{\gamma|\gamma:\mathbf{x}\rightarrow \mathbf{x}_{\rm b}\}}
\int_\gamma \kappa(s) {\rm d} s,
\end{equation}
where $\mathbf{x}_{\rm b}$ is any point of the boundary from which radiation escapes freely.
The surface where radiation decouples from matter is defined as the region where 
$\tau \sim 1$ and it is referred as neutrino surface. If its
curvature is not very pronounced, Eddington approximation applies and a neutrino surface is often referred as the surface where $\tau = 2/3$.

\begin{figure}[!h]
  \begin{center}
    \includegraphics[width=0.5\textwidth]{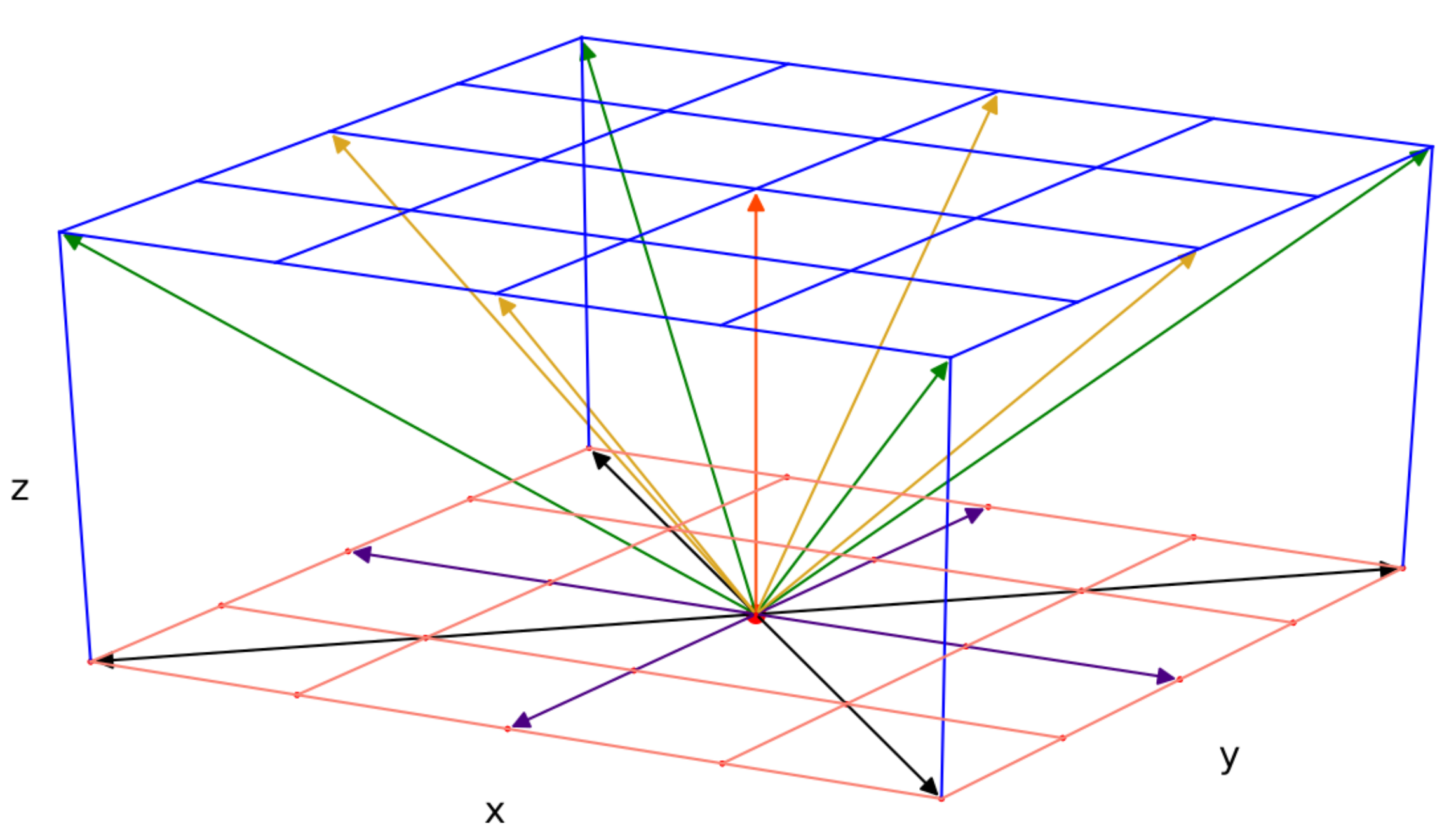}	
  \end{center}
  \caption{Cartoon showing the $17$ directions considered by our scheme in the evaluation of optical depth for neutrinos at the point $\mathbf{x}=(x,y,z)$.}
  \label{fig:17dirs}
\end{figure}

A common approach to effectively evaluate \Eref{eq:tau x} is to select a certain number of direction moving away from $\mathbf{x}$, evaluate $\tau$ along those, and then take the minimum value obtained across all directions.
In the past this has been done already for Newtonian simulation, using cylindrical coordinates and $3$ to $7$ directions (see {\it e.g.}, \cite{Rosswog:2003rv} \cite{Perego:2014fma}). In our scheme, instead, we chose $17$ directions, corresponding to $5$ on-axis directions 
$$ x^+, x^-, y^+, y^-, z^+ \, ,$$ (we disregard all $z^-$ directions, since all simulations considered in this work are performed with $z-$symmetry), $8$ planar diagonals 
$$x^+y^+, x^+y^-, x^-y^+, x^-y^-, x^+z^+, x^-z^+, y^+z^+, y^-z^+ \, ,$$ 
and $4$ full diagonals 
$$x^+y^+z^+, x^+y^-z^+, x^-y^+z^+, x^-y^-z^+ .$$ 
A representation of the directions in a Cartesian grid is given by \Fref{fig:17dirs}.

Even limiting the number of paths to $17$, computing the integrals over the entire simulation domain is usually very expensive in large simulations covering thousands of km in each direction. However, the contribution to the optical depth given by very low-density matter is negligible. For this reason, when performing the integration, we reduce ourselves to a smaller region of the simulated volume centered around the remnant, where we expect to find the neutrino surfaces. In order to select this region, we define the characteristic length of the system $l_C$ as 
\begin{equation}
  l_C = 5 r_{11},
  \label{eq:lc}
\end{equation}
with $r_{11}$ being the average radius of a density isosurface at $\rho_{\rm th} = 2 \times 10^{11}\, \mathrm{g~cm^{-3}}$. In the simulations presented in this work $r_{11} \sim 30 \mathrm{km}$.
To test our approximation, we have computed $\tau$ with $l_C = 5$, $6$, $8\, \, r_{11}$ for a few selected cases, without finding significant differences in the thermodynamics properties of the decoupling region.

\subsection{Neutrino opacities and BNS merger simulations}


Neutrino opacities are provided by weak interaction processes and depends on the local matter properties, on the neutrino flavor and energy, and possibly on the neutrino radiation fields itself. Under the assumption of Nuclear Statistical Equilibrium (NSE), the thermodynamics state of matter is fully identified by its rest mass density, temperature, and electron fraction. In this work we distinguish between three independent neutrino species, namely electron neutrinos ($\nu_e$), electron anti-neutrinos ($\bar{\nu}_e$) and a collective species for $\mu$ and $\tau$ (anti)neutrinos ($\nu_x$). The most relevant neutrino-matter reactions considered in this work are listed in \Tref{tab:leakage}. We consider the corresponding absorptivity $\kappa_{{\rm ab}}$ or scattering opacity $\kappa_{{\rm sc}}$, as they are defined in the Boltzmann transport equation.

On the one hand all processes equally contribute to the neutrino opacity relevant for the diffusion process. We thus define the diffusion opacity $\kappa_{\rm diff}$ as 
\begin{equation}
 \kappa_{{\rm diff}} = 
 \sum\limits_{r} \kappa_{{\rm ab},r} + \sum\limits_{s} \kappa_{{\rm sc},s} \, ,
 \label{eq:kappa_diff}
\end{equation}
where the indexes $r$ and $s$ run over all the considered absorption and scattering reactions, respectively.
On the other hand only a subset of reactions are effective in keeping the neutrino field 
in thermal and weak equilibrium with the plasma. We define an equilibrium opacity as
the geometrical mean between the diffusion opacity and the opacity only due to
absorption processes (see {\it e.g.} \cite{Shapiro:1983du} or \cite{Raffelt:2001kv} for analogous expressions):
\begin{equation}
 \kappa_{{\rm eq}} = \sqrt{ \left( \sum \limits_{r} \kappa_{{\rm ab},r} \right) \kappa_{{\rm diff}}}  \, .
 \label{eq:kappa_eq}
\end{equation}
These two kinds of opacities provide two different optical depths, $\tau_{\rm diff}$ 
and $\tau_{\rm eq}$, and two different kinds of neutrino surfaces.
The diffusion surfaces mark the transition from the semi-transparent to the optically thin (free-streaming) regime.
They can be identified as the last interaction surfaces, meaning that outside them neutrinos are likely not to interact anymore
with matter, regardless of the nature of the reaction.
On the other hand, the equilibrium surfaces identify the conditions at which neutrino radiation decouples from the background 
medium, in the sense that freeze-out from thermal and weak equilibrium occurs. This equilibrium is guaranteed by inelastic reactions
that produce and absorb neutrinos. Since the latter form a subset of all reactions, $\kappa_{\rm diff} \geq \kappa_{\rm eq}$
and the equilibrium surfaces lie always inside the diffusion ones. In particular, if quasi-elastic scattering is efficient 
in providing opacity beyond the equilibrium surfaces, a scattering atmosphere, where neutrinos diffuse far from local equilibrium,
can form.

To compute energy-dependent neutrino opacities for neutrino absorption on nucleons and scattering off nucleons and nuclei we use the publicly available library \texttt{NuLib} \cite{OConnor:Thesis:2012}. In particular, we rely on expressions for the transport opacities and we include weak magnetism corrections \cite{Horowitz:2001xf}, ion-ion correlations, form factor correction \cite{Tubbs:1975jx,Burrows:1981zz}, and electron polarization correction \cite{Leinson:1997ns} according to \cite{Burrows:2004vq}. Differently from the standard code version, we do not include the effect of stimulated absorption in our calculations because it provides unphysically high
$\bar{\nu}_e$ opacities in cold, low dense matter, well below $\rho_{\rm th}$.
For the inverse nucleon-nucleon bremsstrahlung and the neutrino-antineutrino pair annihilation we compute the reaction kernels according to \cite{Hannestad:1997gc} and \cite{Bruenn:1985en,Mezzacappa:1993gn}, respectively. These kernels depend on the species and energies of both incoming neutrinos. Computations of the corresponding opacities would require the knowledge of the detailed neutrino distribution functions, since neutrinos are not only the colliding,  but also the target particles. Since this information is in general not available, the absorption opacity for each neutrino species and energy is computed assuming the target neutrinos to be in equilibrium with matter (for example, in the case of $\nu_e + \bar{\nu}_e \rightarrow e^+ + e^-$, we fix the $\nu_e$ energy and we integrate over an equilibrium Fermi-Dirac distribution function for $\bar{\nu}_e$ ).
While this hypothesis is not correct in optically thin conditions, it is well verified in neutrino trapped regions. Since in this work we explore the intermediate, semi-transparent regime, the validity of this approach is a priori uncertain. In general, due to the pair nature of the reaction, the corresponding opacity drops quickly outside the neutrino surfaces even assuming equilibrium conditions for the target particles. Thus, even if not fully correct, their contributions to the absorption opacity in optically thin conditions in \Eref{eq:tau x} is small and we do not expect a significant change in the calculations of the optical depths and in the determination of the neutrino surfaces. Moreover, these processes are relevant in the case of $\nu_x$'s. Due to the presence of an extended scattering atmosphere, we expect heavy lepton neutrinos to form a trapped gas also when pair processes become inefficient in keeping the radiation in equilibrium with matter. Thus, even if not accurate, we consider our approximation reasonable in the determination of the location of the neutrino surfaces for all neutrino species.

The thermodynamics conditions of matter used as input for the neutrino opacity calculations
are taken from the output of (3+1)D numerical relativity simulations of BNS mergers.
We consider equal mass binaries with 
$M_{\rm NS} = 1.364~M_{\odot}$ 
(corresponding to a chirp mass of $1.188~M_{\odot}$ compatible with GW170817) and we 
make use of two different finite temperature, composition 
dependent EOSs: DD2 \cite{Typel:2009sy,Hempel:2009mc} and SLy4 \cite{daSilvaSchneider:2017jpg}. 
These tables assume NSE to determine the nuclear composition. 
We produce two distinct opacity tables for the two different simulations, using the appropriate EOS table as input.
We solve the general relativistic initial data problem to produce
irrotational BNS configurations in quasi-circular orbit using 
the \texttt{Lorene} pseudo-spectral code \cite{Gourgoulhon:2000nn}.
The initial separation between the NS is set to $45\, {\rm km}$, corresponding
to $\sim2{-}3$ orbits before merger.
Low temperature ($T\lesssim 0.1$~MeV), $\nu$-less weak equilibrium slices of the EOS 
are employed to construct the initial data. To correct for the presence of photons
at low density we subtract their pressure contribution from the cold slices.

The (3+1)D evolutions are performed with the \texttt{WhiskyTHC} code
\cite{Radice:2012cu,Radice:2013hxh,Radice:2013xpa,Radice:2015nva}, complemented by a leakage scheme to account for compositional and energy changes in the matter 
due to weak reactions involving $\nu_e$, $\bar{\nu}_e$, and $\nu_x$. Free-streaming neutrinos are emitted at an average energy and then evolved according to the M0 scheme introduced in \cite{Radice:2016dwd,Radice:2018pdn}.
All the technical details for these simulations are given
in \cite{Radice:2018xqa,Perego:2019adq} which we refer to also for the employed grid setup. 
A detailed discussion of the thermodynamics properties of the matter
during merger can be found in \cite{Perego:2019adq}.
The DD2 simulation is the same discussed
in \cite{Perego:2019adq}, the SLy4 simulation is presented here for the first time.
The domain covered by our simulation is a cube of size length 3,024~km (assuming mirror symmetry along $z$) whose center is at the center of mass of the binary. 
It is resolved by a Cartesian grid hierarchy composed of 7
2:1 nested refinement levels. The finest refinement level, covering both NSs
during the inspiral and the central remnant after merger, has a 
resolution of $h \simeq 185\, {\rm m}$.
Since our optical depth calculation scheme works on uniform grids, we interpolate the simulation results onto one uniform grid of resolution $h' \sim 0.74 \mathrm{km}$ representing our standard resolution. 
In Appendix~\ref{app:convergence} we study the sensitivity of our results by running both higher and lower resolution postprocessing analysis, and we find no significant differences.
We repeat our analysis also using a different optical depth computation scheme, and we find again no significant difference, as visible in Appendix~\ref{sec:sccomp}.
Due to the approximate nature of our neutrino transport scheme, the results we present in this paper carry a certain degree of uncertainty. Comparison with simulations performed with different neutrino schemes ({\it e.g.} \cite{Perego:2017xth}, however, did not reveal significant differences in the density and temperature profiles. On the other hand, the electron fraction in the remnant is more sensitive to the neutrino transport, especially in the density interval where high energy neutrino decoupling occurs.


The two binary models have 
been selected in order to span the different outcomes of a BNS: the
DD2 EOS model results in a massive NS remnant, which lives longer
than the extent of the simulation, while the SLy4 EOS remnant
collapses to BH around $\sim12\, \mathrm{ms}$ after merger. 
We chose three time snapshots at which to evaluate the
optical depth that were reached by both simulations.
The timesteps correspond to $1$, $10$, and $20\, \mathrm{ms}$ after merger time (identified as the time corresponding to the maximum in the strain of the $\ell=m=2$ mode of the gravitational wave signal).
Three dimensional profiles of the density, temperature, 
electron fraction, and spatial metric tensor were extracted
for all the three snapshots and from both simulations. 

\begin{table}
\caption{Weak reactions providing the neutrino opacities used in this work 
and references for their implementation.
$\nu \in \{\nu_e, \bar{\nu}_e, \nu_{x}\}$
denotes a neutrino species with $\nu_{x}$ referring to any heavy-lepton 
neutrino species. 
$N \in \{n, p\}$ denotes a nucleon, $A$ a generic nucleus (including $\alpha$ particles), $e^{\pm}$ electrons and positrons.} 
\label{tab:leakage}
\begin{center}
  \begin{tabular}{ll}
\hline\hline
Reaction & Ref. \\ 
\hline
$\nu_e + n \rightarrow p + e^-$                & \cite{Burrows:2004vq,Horowitz:2001xf} \\
$\bar{\nu}_{e} + p \rightarrow n + e^+$        & \cite{Burrows:2004vq,Horowitz:2001xf} \\
$\nu + \bar{\nu} \rightarrow  e^+ + e^- $     & \cite{Bruenn:1985en,Mezzacappa:1993gn} \\
$N + N + \nu + \bar{\nu} \rightarrow N  + N$  & \cite{Hannestad:1997gc} \\
$\nu + N \rightarrow \nu + N$                 & \cite{Burrows:2004vq} \\
$\nu + A \rightarrow \nu + A$                 & \cite{Burrows:2004vq} \\
\hline\hline
\end{tabular}
\end{center}
\end{table}

\subsection{Analysis strategy}
\label{subs: analysis strategy}

Neutrino opacities have a significant dependence on the energy of the incoming neutrino.
A crucial ingredient is thus the determination of the neutrino energy at which the 
opacity entering \Eref{eq:tau x} should be evaluated.

Numerical simulations of BNS mergers including neutrino radiation
provide values for the average energies of the neutrinos escaping to infinity.
Despite the large variety of approximated schemes employed in these analysis, reported
values usually agree within 10-20\%~\cite{Foucart:2015vpa,Perego:2015agy,Foucart:2016rxm}

As a first approach, we fix the neutrino energy to the following set of values compatible with those results: $\langle E_{\nu_e} \rangle \approx 9.34~{\rm MeV}$, $\langle E_{\bar{\nu}_e} \rangle \approx 15.16~{\rm MeV}$,  and $\langle E_{\nu_x} \rangle \approx 23.98~{\rm MeV}$, 
and we compute spectral optical depths and neutrino surfaces for them. This approach assumes that the neutrino spectrum at infinity is mainly determined by the spectrum emerging from the neutrino surfaces. We determine the thermodynamics conditions of matter (density, temperature and
electron fraction) at the neutrino surfaces and we use them to characterize
the typical conditions at which the largest fraction of neutrinos decouples from matter.
In our discretized domain, we identify the neutrino surface as the region where $0.5 \leq \tau \leq 0.85$.
For each extracted quantity $q$ we compute the volumetric mean $q_{\rm mean}$ and to give an estimate of its distribution around the mean value we compute the corresponding standard deviation as:
\begin{equation}
 \sigma_q = \frac{1}{N} \sum_{i=1}^{N} \sqrt{ \left( q_i-q_\mathrm{mean} \right)^2  } \, ,
\end{equation}
where the index $i$ runs over all $N$ cells belonging to the neutrino surface. Due to the actual distribution of conditions, we choose $q$
to be: $\log_{10}\left( {\rho}~[{\rm g~cm^{-3}}] \right)$, $\log_{10}{\left( T~[{\rm MeV}] \right)}$ and $Y_e$.

As a second approach, we compute the optical depths and the corresponding neutrino surfaces for a large set of neutrino energies between $3$ and $88.67$ MeV, and analyze the decoupling conditions as a function of
the neutrino energy. This energy range covers the most relevant part of the neutrino spectra emerging from a NS merger.
Within this analysis we will investigate how different the decoupling thermodynamics conditions are for neutrinos of different energies, regardless of their importance in the emerging spectrum.

Finally, we compute the optical depths using energy-integrated
opacities. This case is relevant since the few compact
binary merger simulations with the M1 transport
schemes for neutrinos in the literature are performed assuming a gray approximation
~\cite{Foucart:2015vpa,Foucart:2016rxm}. The latter consists in prescribing a certain functional
form for the neutrino distribution function, thus removing the $3N$ degrees of freedom represented by $N$ energy groups for each species.
When neutrinos are coupled to matter, weak and thermal equilibrium drives their distributions to an isotropic Fermi-Dirac distribution:
\begin{equation}
f_{\nu} (E_\nu) = \left( e^{(E_\nu - \mu_{\nu})/k_b T} +1 \right)^{-1} \, ,
\label{eq:fermi}
\end{equation}
where $E_\nu$ is the neutrino energy, $\mu_\nu$ the neutrino chemical potential, 
$k_b$ is the Boltzmann constant, and $T$ is the temperature of the fluid at absorption (or scattering) point. We evaluate the neutrino chemical potentials at equilibrium as 
\begin{eqnarray}
\mu_{\nu_e} &=& \mu_p +\mu_e -\mu_n,\nonumber\\
\mu_{\bar{\nu}_e} &=& - \mu_{\nu_e},\\
\mu_{\nu_x} &=& 0\, , \nonumber
\end{eqnarray}
where $\nu_n$, $\nu_p$ and $\nu_e$ are the relativistic neutron, proton and electron chemical potentials, respectively.
We introduce a spectrum-averaged opacity defined as:
\begin{equation}
 \label{eq:average_opacity}
 \tilde{\kappa}_{\nu} = \frac{\int_0^{\infty} f_{\nu} (E) \kappa_{\nu}(E) E^2 dE}{\int_0^{\infty} f_{\nu} (E) E^2 dE } \, .
\end{equation}
By calculating this average, we compute the typical opacity locally experienced by neutrinos at equilibrium and diffusing from optically thick towards optically thin regions.
Since this approach assumes that neutrinos are in equilibrium condition with matter, it is more appropriate for neutrinos in optically thick conditions and more in line with the equilibrium optical depth, $\tau_{\rm eq}$.
The average used in this work, \Eref{eq:average_opacity}, is slightly different from the average computed in some gray neutrino transport schemes, {\it e.g.} \cite{Foucart:2015vpa}, where the factor $E^2$ is replaced by $E^3$. In the latter approach, more emphasis is put on the energy transport ($I_{\nu} \propto E^3$) rather than on the particle transport. We do not expect a qualitative difference between the two approaches, even if the energy transport opacities are usually larger than the particle transport ones.


\section{Results}
\label{sec:results}

\subsection{Opacity variability}
\label{subs:opvar}

\begin{figure*}[!t]
  \begin{center}
    \includegraphics[width=\textwidth]{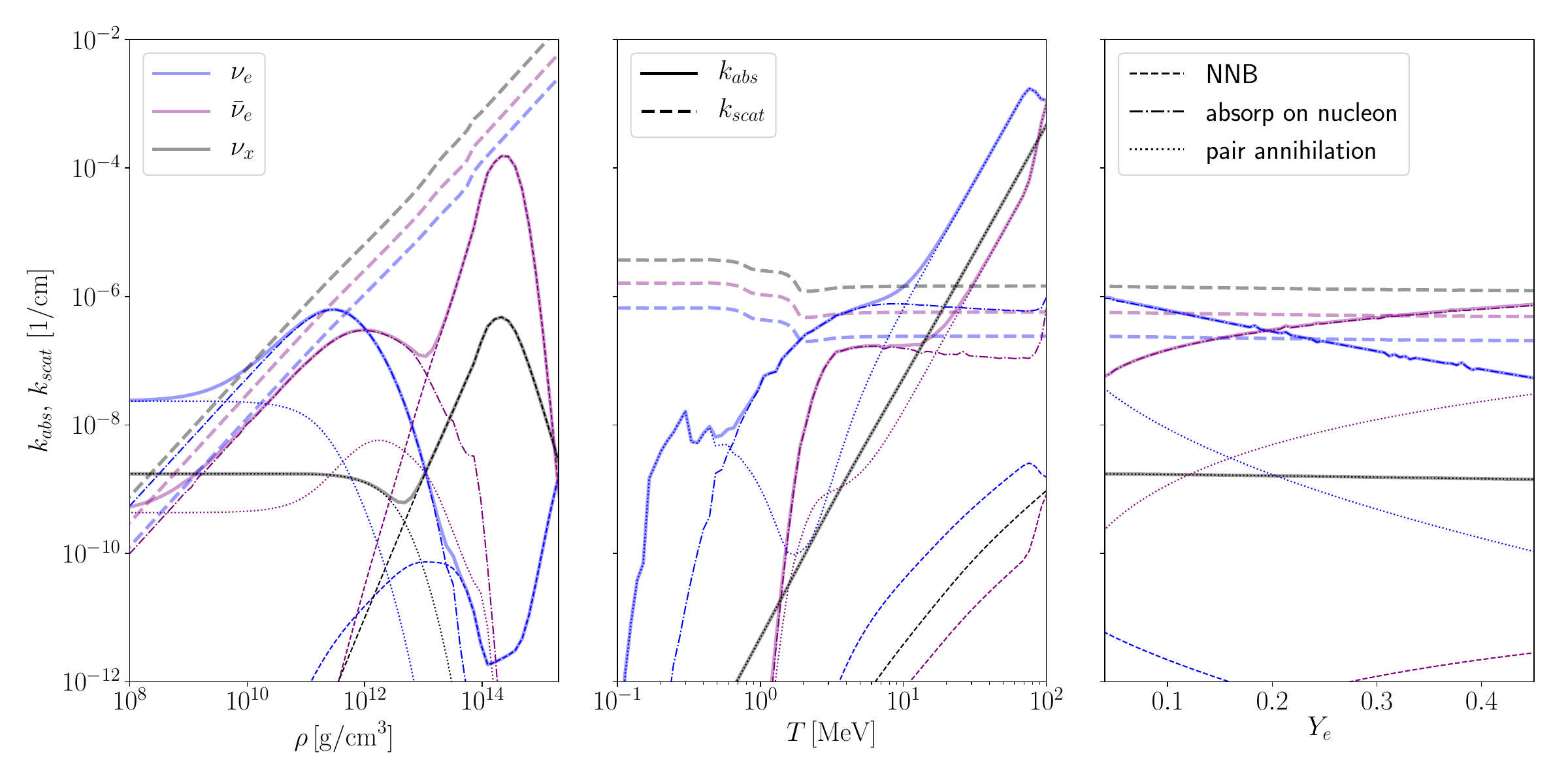}	
  \end{center}
  \caption{Opacity contributes from absorption ($k_{\rm abs}$) and scattering ($k_{\rm scat}$) processes for the three considered neutrino species, each evaluated at its average energy $\langle E_{\nu} \rangle$, as function of the main thermodynamics variables ($\rho$, left; $T$, middle; $Y_e$,left) for the DD2 EOS. In each panel the other two variables are set to values close to what we expect in the region where neutrinos decouple ($\rho_{\rm th} = 2 \times 10^{11}\mathrm{g~cm^{-3}}$, $T_{\rm th} = 4 \mathrm{MeV}$, $Y_{e,\, {\rm th}} = 0.1$). The thick lines represent the full opacities, while the thin ones are the single contributes to the absorption opacity, including inverse nucleon nucleon bremsstrahlung (NNB), $\nu$-$\bar{\nu}$ pair annihilation, and absorption on free nucleons.}
  \label{fig:opac_dd2}
\end{figure*}


We start our analysis by exploring the opacities for the different neutrino species. 
In \Fref{fig:opac_dd2}, we present the dependency of the absorption and scattering opacities ($k_{\rm abs}$ and $k_{\rm scat}$, represented by solid and dashed lines, respectively) on rest mass density, temperature and electron fraction of the medium for the three neutrino species. We consider thermodynamics conditions close to the regions where we expect mean energy neutrinos to dynamically decouple from matter ($\rho_{\rm th} = 2\times 10^{11}~{\rm g~cm^{-3}}$, $T_{\rm th}= 4~{\rm MeV}$ and $Y_{e,{\rm th}}=0.1$, see for example \cite{Perego:2014fma}). In particular we vary one of the three variables and keep the other two fixed. For concreteness, we fix the energy of the incoming neutrinos to the mean energies $\langle E_{\nu} \rangle $ reported in \Sref{subs: analysis strategy} and we use the DD2 EOS to compute nuclear properties, abundances and chemical potentials for matter in NSE.

Due to the dominant role of quasi-elastic scattering off free nucleons, scattering opacities present a very similar behavior among the three neutrino species. Since in very good approximation $\kappa_{\rm scat} \propto n_{\rm b} E_{\nu}^2$ (with $n_{\rm b}$ being the free baryon particle density), $k_{\rm scat}$ is approximately proportional to $\rho$, while the larger (smaller) scattering opacity observed for $\nu_x$ ($\nu_e$) results from the larger (smaller) neutrino mean energy.
Due to the quasi-elastic nature of the dominant scattering process and to the isotropic spatial distribution of all interacting particles, $k_{\rm scat}$ is rather insensitive to variations in temperature and electron fraction. Weak magnetism, recoil and nucleon form factors introduce differences in the cross sections possibly sensitive to the neutron-to-proton content, but on a scale much smaller than the variation induced by the change in nucleon number density and in the neutrino energy. The increase of $k_{\rm scat}$ at low temperatures ($T \lesssim 1~{\rm MeV}$) visible in the central panel is related with the appearance of bound nuclei, instead of free baryons, that provide a more effective coherent scattering, see {\it e.g.} \cite{Bruenn:1985en,Burrows:2004vq}.

In the case of $k_{\rm abs}$, the dependence on the thermodynamics conditions is more complex, due to the presence of several processes characterized by different target particles. 
In \Fref{fig:opac_dd2}, we present also the different contributions to the absorption opacity for the three neutrino species, including the inverse nucleon-nucleon bremsstrahlung, the absorption on free nucleons, and the annihilation of neutrino-antineutrino pairs. As visible in the left panel, for densities lower than $\sim 10^{12}{\rm g~cm^{-3}}$ the neutron richness and the larger $Q$-value favor $\nu_e$ captures on neutrons with respect to $\bar{\nu}_e$ captures on protons\footnote{For $\rho \lesssim 10^{10}{\rm g~cm^{-3}}$, the dominant role of neutrino pair annihilation is a spurious effect due to the assumption of having a neutrino gas even at low densities. We notice that, even for a moderately high temperature, the values for the opacities are relatively low and thus they do not impact on the determination of the decoupling conditions.}. For larger densities, Pauli blocking for the more degenerate final state particles drastically reduces the opacity due to charged current absorption processes. For densities in excess of $10^{13}{\rm g~cm^{-3}}$, assuming again temperatures around a few MeV, the inverse nucleon-nucleon bremsstrahlung becomes the dominant absorption process. Depending on the neutrino chemical potentials at equilibrium, the abundance of the target neutrinos can significantly change. For example, for the explored conditions $\bar{\nu}_e$ are suppressed by degeneracy at high densities and this translates in a much smaller inverse bremsstrahlung opacity for $\nu_e$ compared to $\nu_x$ and 
$\bar{\nu}_e$.
Variations in temperature have also a significant effect ok $k_{\rm abs}$, as visible in the central panel of \Fref{fig:opac_dd2}. The final state blocking effect that characterizes charged current absorption reactions is reduced once the temperature increases between one and a few MeV, while for temperatures in excess of $\sim 10$~MeV pair processes of thermal origin (like $\nu$-$\bar{\nu}$ annihilation) dominate. 
In particular, the corresponding inverse mean free path is roughly proportional to the $Y_{\nu} Y_{\bar{\nu} \sigma_{\nu+\bar{\nu}}}$, which for thermally produced neutrinos translate in a strong dependence on the matter temperature, $\propto T^8$.

Changes in the electron fraction also introduce significant dependencies on the $k_{\rm abs}$ contributions. They are ultimately due to changes in the number densities of the target particles in the case of charged current absorption processes, and to variations in the electron chemical potential for neutrino pair annihilation.

\subsection{Conditions at the decoupling surfaces for mean neutrino energies}
\label{subs:edep}

\begin{figure*}[!t]
	\begin{center}  
		\includegraphics[width=\textwidth]{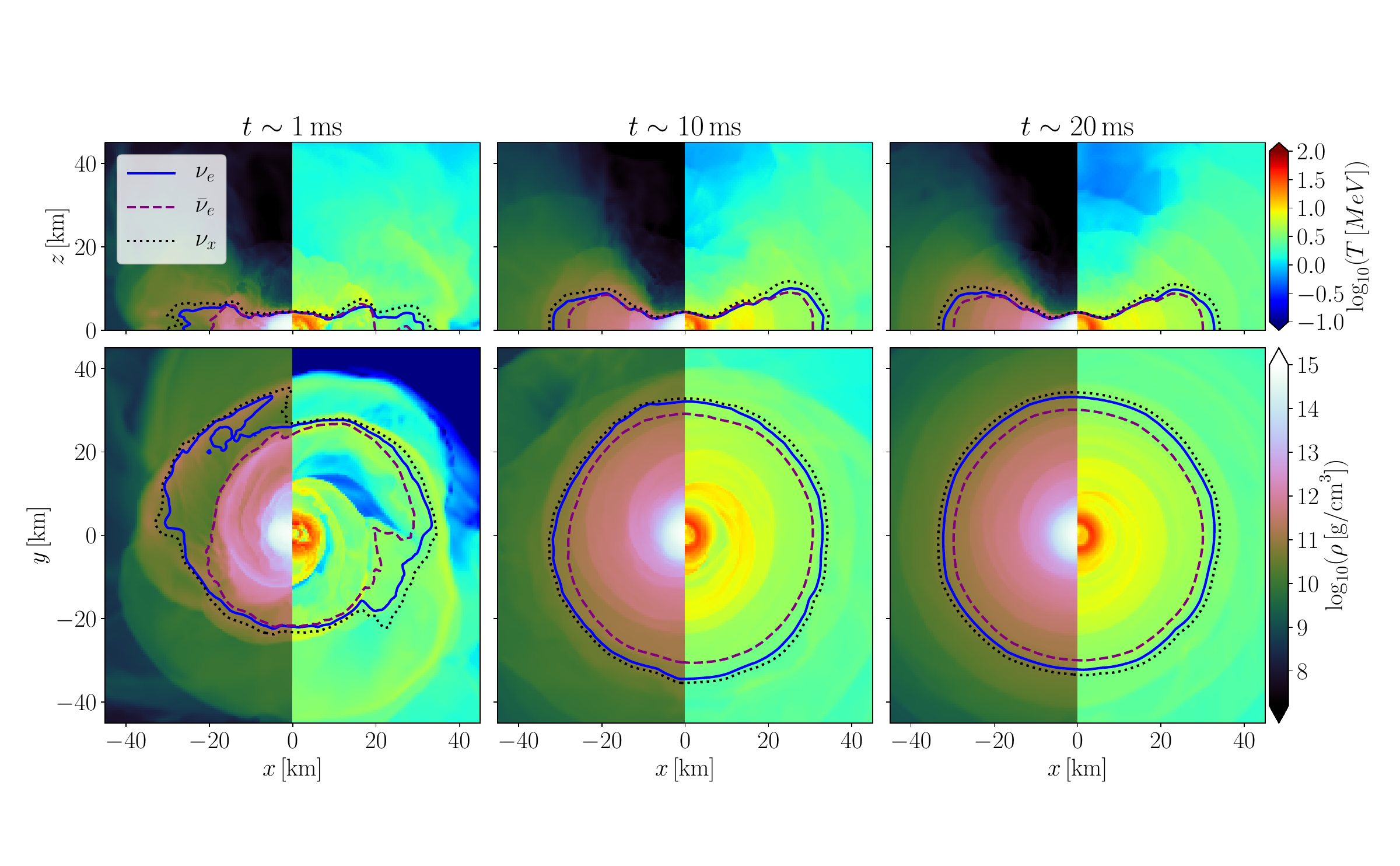}
	\end{center}
	\caption{Evolution of diffusion neutrino surfaces for the mean neutrino energies 
	      at different stages of
          the post-merger of the DD2 model.
          The color code represents rest mass density (left side) and temperature (right side) in logarithmic scale in both meridional (top row) and equatorial (bottom row) plane. The three snapshots are taken at the closest iteration to respectively $1$ (left column), $10$ (middle column) and $20$ (right column) $\mathrm{ms}$ after merger. Diffusion surfaces for neutrinos of the three considered species are highlighted by solid blue ($\nu_e$), dashed magenta ($\bar{\nu}_e$) and dotted black ($\nu_x$) contours.}
	\label{fig:dd2_evol}
\end{figure*}

\begin{figure*}[!t]
	\begin{center}  
		\includegraphics[width=\textwidth]{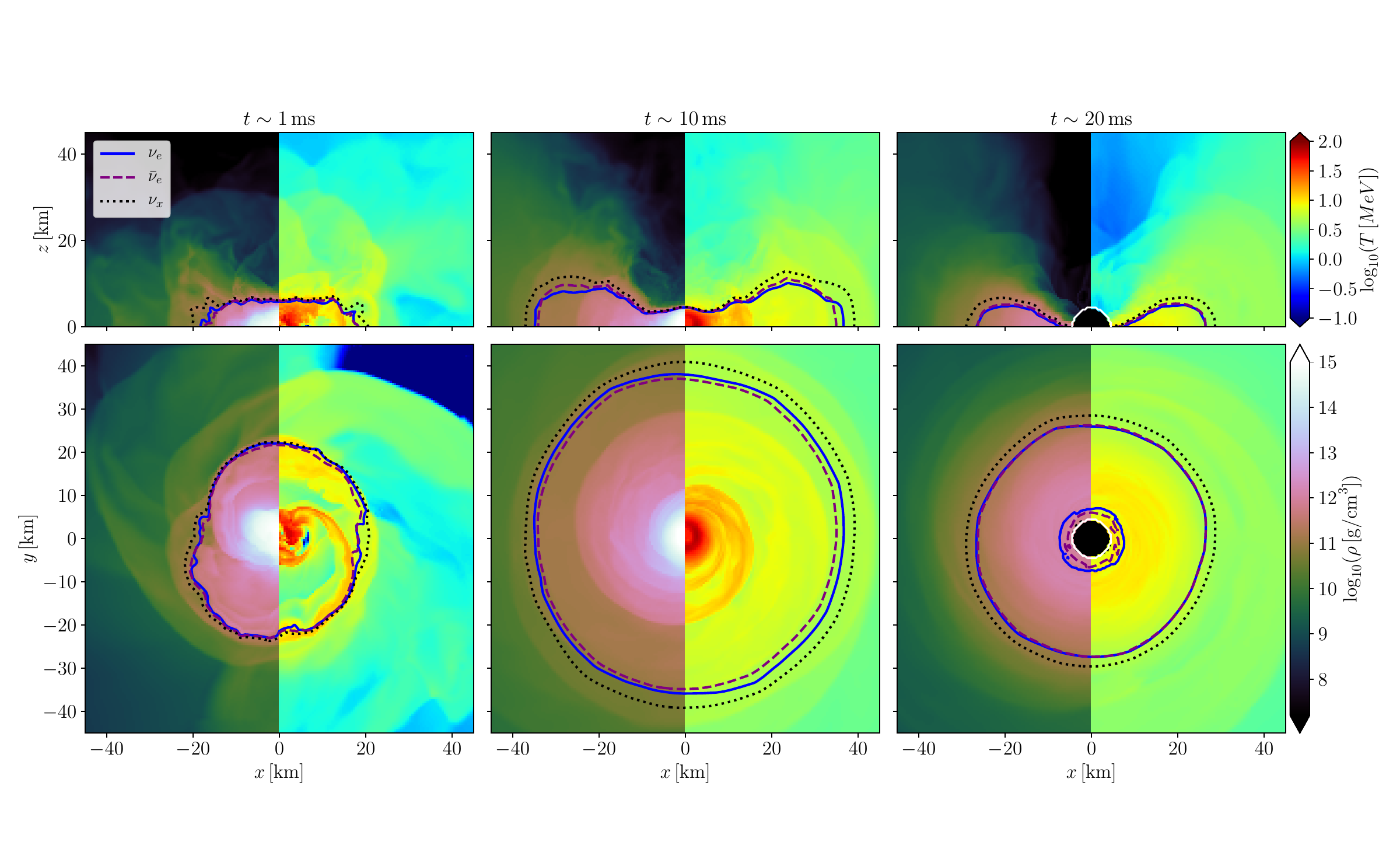}
	\end{center}
	\caption{Same as \Fref{fig:dd2_evol} for the diffusion neutrino surfaces for the mean neutrino energies at different stages of
          the post-merger of the SLy4 model.}
	\label{fig:sly4_evol}
\end{figure*}

 \begin{table*}[!t]
  \caption{In this table we present the mean values of density $\rho$, temperature $T$ and electron fraction $Y_e$ in the regions with opacity $0.5\leq\tau_{s}\leq0.85$ using \Eref{eq:kappa_diff} for both models at different stages of the postmerger evolution. In the Sly4 case, at $t\sim20$ ms, the remnant already collapsed to BH. Densities are given as $\log_{10}(\rho\, [\mathrm{g cm}^{-3}])$ and temperatures as $\log_{10}(T\, [\mathrm{MeV}])$. For each quantity we also give its variance in the distribution.}
    \begin{tabular}{|l|ccc|ccc|}
     \hline
      & & DD2 & & & SLy4 & \\
     \hline
      & $t\sim1$ ms & $t\sim10$ ms & $t\sim20$ ms &
        $t\sim1$ ms & $t\sim10$ ms & $t\sim20$ ms \\
     \hline
     $\log_{10}(\rho^{\nu_e})$ & 
        $10.89\pm0.33$ & $10.94\pm0.23$ & $10.94\pm0.23$ &
        $11.36\pm0.22$ & $10.97\pm0.22$ & $11.33\pm0.24$ \\
     $\log_{10}(T^{\nu_e})$ & 
        $0.49\pm0.14$ & $0.58\pm0.09$ & $0.58\pm0.07$ &
        $0.81\pm0.20$ & $0.75\pm0.06$ & $0.75\pm0.07$ \\
     $Y_e^{\nu_e}$ &
        $0.14\pm0.05$ & $0.14\pm0.02$ & $0.14\pm0.01$ &
        $0.20\pm0.08$ & $0.25\pm0.02$ & $0.18\pm0.01$ \\
     \hline
     $\log_{10}(\rho^{\bar{\nu}_e})$ & 
        $11.07\pm0.28$ & $11.07\pm0.17$ & $11.07\pm0.19$ &
        $11.35\pm0.20$ & $10.97\pm0.18$ & $11.30\pm0.22$ \\
     $\log_{10}(T^{\bar{\nu}_e})$ & 
        $0.52\pm0.16$ & $0.62\pm0.07$ & $0.61\pm0.06$ &
        $0.82\pm0.17$ & $0.75\pm0.06$ & $0.74\pm0.06$ \\
     $Y_e^{\bar{\nu}_e}$ &
        $0.13\pm0.05$ & $0.15\pm0.02$ & $0.14\pm0.01$ &
        $0.21\pm0.08$ & $0.25\pm0.02$ & $0.18\pm0.01$ \\
	 \hline
     $\log_{10}(\rho^{\nu_x})$ & 
        $10.76\pm0.25$ & $10.80\pm0.18$ & $10.80\pm0.19$ &
        $11.20\pm0.19$ & $10.82\pm0.18$ & $11.11\pm0.21$ \\
     $\log_{10}(T^{\nu_x})$ & 
        $0.48\pm0.14$ & $0.55\pm0.09$ & $0.55\pm0.06$ &
        $0.82\pm0.17$ & $0.71\pm0.05$ & $0.69\pm0.05$ \\
     $Y_e^{\nu_x}$ &
        $0.14\pm0.05$ & $0.14\pm0.02$ & $0.13\pm0.01$ &
        $0.24\pm0.09$ & $0.24\pm0.02$ & $0.18\pm0.02$ \\
    \hline
    \end{tabular}
    \centering
  \label{tab:stau_ev}
\end{table*}

\begin{figure*}[!t]
	\begin{center}  
		\includegraphics[width=\textwidth]{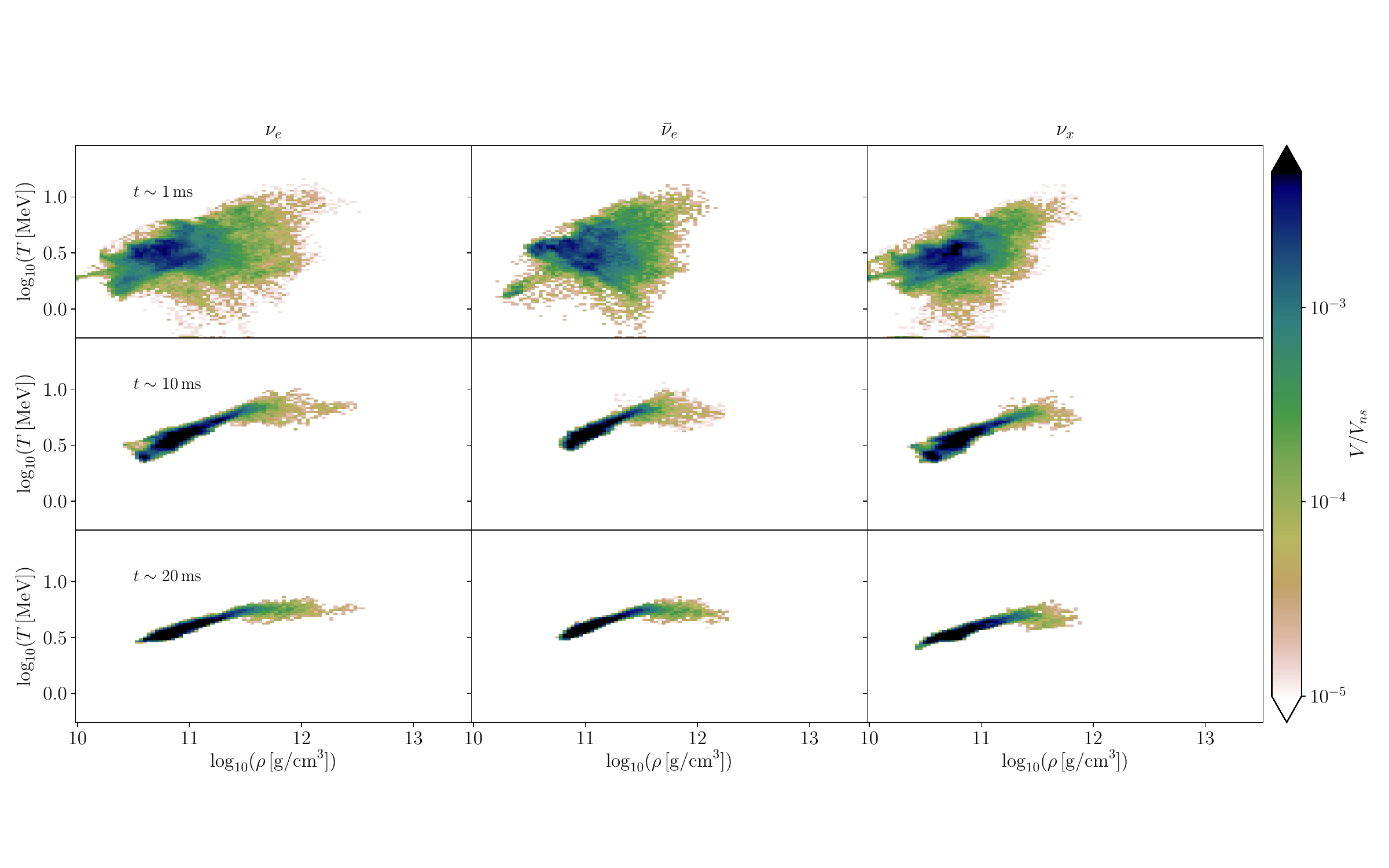}
	\end{center}
	\caption{Histograms of the occurrence of density and temperature conditions for matter in the neutrino decoupling region around the diffusion neutrino surfaces ($0.5< \tau_{\rm diff} <0.85$) for the DD2 model. Color coded is the occurrence volume, normalized to the total volume of the neutrino surface ($V_{\rm ns}$). From left to right we show the distributions for the three considered neutrino species, while each row represents a different time after merger ($1$, $10$ and $20\, \mathrm{ms}$ respectively).}
	\label{fig:dd2_T_mdist_diff}
\end{figure*}

We move to the analysis of the neutrino surfaces and the corresponding thermodynamics conditions obtained for the average neutrino energies $\langle E_{\nu} \rangle$. 

\subsubsection{Diffusion surfaces}

In \Fref{fig:dd2_evol} and \Fref{fig:sly4_evol} we present the contours of the diffusion surfaces ($\tau_{\rm diff} = 2/3$) at three different snapshots for the DD2 and SLy4 models, respectively. For each snapshot we show the rest mass density and temperature (respectively on the left and right half of each panel) in the meridional (top row) and equatorial (bottom row) planes.
While around $10~{\rm ms}$ after merger the system has already reached an almost axisymmetric configuration, in the first snapshot ($\sim 1~{\rm ms}$ after merger) the density and temperature profiles reveal a still highly dynamical evolution. The latter is characterized by the presence of spiral arms that rotate and shock matter in the mantle forming around the merging cores. 

Due to the quadratic dependence on the neutrino energy of the elastic scattering off nucleons, the neutrino surfaces of heavy flavor neutrinos are 
the most extended. At the same time, despite the lowest $\nu_e$ mean energy, the more abundant free neutrons provide larger opacities to $\nu_e$ than to $\bar{\nu}_e$, {\it i.e.}
$(1-Y_e) \langle E_{\nu_e} \rangle^2 \gtrsim Y_e \langle E_{\bar{\nu}_e} \rangle^2$. Thus, the average $\bar{\nu}_e$'s start to stream freely from (marginally) smaller radii.
The last snapshots in \Fref{fig:dd2_evol} and \Fref{fig:sly4_evol} show two different remnants: while
the DD2 massive NS still survives, in the SLy4 simulation a BH has formed. In \Fref{fig:sly4_evol} we highlight the
apparent horizon (AH) with a white contour and black fill. As expected, when a BH has formed the neutrino surface shape changes. In the case of $\nu_e$ and $\bar{\nu}_e$ an outer and inner contour are both present, resulting in a toroidal-shaped neutrino surfaces. The heavy flavor neutrinos instead only show an outer contour, while the inner contour was swallowed inside the AH\footnote{We note that the region within the AH is not considered in our evaluation of hydrodynamics properties in the transient region.}.
In the DD2 case, where the massive NS is still surviving, the degree of axisymmetry and the disk compactness increase moving from $10$ to $20\, \mathrm{ms}$.
In terms of neutrino surfaces, we notice that for more compact disks, corresponding to late time configurations or to the softer SLy4 EOS, the distance between the outer electron neutrino and antineutrino surfaces decrease, due to the steeper density gradients.

The hierarchy observed in the neutrino surface locations is also visible in the mean densities and temperatures recorded at the diffusion surfaces and summarized in \Tref{tab:stau_ev}. 
The mean values obtained for $\log_{10}{\left( \rho \right)}$ for each model and at any snapshots are well representative, since their spreads are rather small compared to the mean values (a few percent) and to the large range over which the matter density varies.
Moreover, for each neutrino species they are rather insensitive to the model and to the time evolution (at least, within the uncertainty represented by their spreads).
More variability is instead observed in the temperature distributions: the mean temperatures are significantly larger for the SLy4 model (5-6 MeV) than for the DD2 model (3-4 MeV), as expected due to the larger temperatures that characterize soft EOS remnants. Moreover, in both cases the spreads around the means are of the order of 1/3 of the mean value just after merger and reduce down to $\sim 1$~MeV at later time (corresponding to the 10 and 20 ms snapshots).
This is a consequence of the evolution of matter distribution in the density-temperature plane, see \cite{Perego:2019adq} and the histograms below.
In the early post-merger the gradient of temperature around the diffusion surfaces can be quite steep, due to the hot tidal tails extending from the remnant into the forming accretion disk. This feature is present in both EOS models, even though in the DD2 model the tidal tails are more defined and develop stronger gradients with respect to the SLy4 model. This leads to a higher variability in the estimation of the mean temperature in the transient region at early stages of the remnant evolution. At $\sim10\, \mathrm{ms}$ after merger, the outflow of matter from the remnant injects enough hot and dense matter into the disk such that the transient region for the diffusion surfaces is pushed further out and the tidal tails become less prominent.
For times larger than 10~ms the remnants are more axisymmetric, and the hot tidal tails transfer heat to the material in the transient region more uniformly, resulting in almost stationary mean temperatures and lower spreads. This is especially clear in the SLy4 model, where the temperature spread in the decoupling volume is halved after $9\, \mathrm{ms}$.
A similar trend is also observed for $Y_e$ at the diffusion neutrino surfaces: remnants obtained by softer EOSs are characterized by larger $Y_e$ at the decoupling surfaces, while the more axisymmetric structure obtained at late times ($\gtrsim 10$~ms after merger) reduces significantly the associated spreads.

%
%

To reveal possible correlations between the decoupling rest mass density and temperature, in \Fref{fig:dd2_T_mdist_diff} we present histograms describing how the volume around the neutrinosphere is distributed
in terms of the matter density and temperature. We consider each of the three snapshots of the DD2 model. More specifically, we show the volumetric amount of matter at specific $(\rho,T)$ conditions normalized to the total volume of the neutrinosphere region $V_{\rm ns}$.
These histograms display two important features. First, the initial large spreads in density and in temperature are initially weakly correlated, but the latter (for a fixed density) diminishes noticeably when moving further away from merger time. Second, at late times there is a clear correlation between the density and temperature conditions. In particular, the bulk of mass shifts towards lower densities and temperatures in the first $10\, \mathrm{ms}$, while maintaining a noticeable spread in temperature in the high density regions. This feature persists for the DD2 model even at $20\, \mathrm{ms}$, while in the SLy4 case (not shown in the figure) the average temperature (at any given density) steadily increases with density, while the temperature spread becomes narrower. This difference is due to the presence of a massive remnant in the DD2 model, whose outflow in the inner disk region keeps a higher variability in density and temperature profiles in the high density part of the decoupling region.
The correlation between the matter density and the temperature reflects the dominant dependency on $\rho$ both for $k_{\rm abs}$ and $k_{\rm sc}$ (see \Sref{subs:opvar}), and trace back to the correlation between density and temperature inside
the remnant, which becomes narrower for increasing time after merger \cite{Perego:2019adq}.


\begin{figure*}[!t]
	\begin{center}  
		\includegraphics[width=\textwidth]{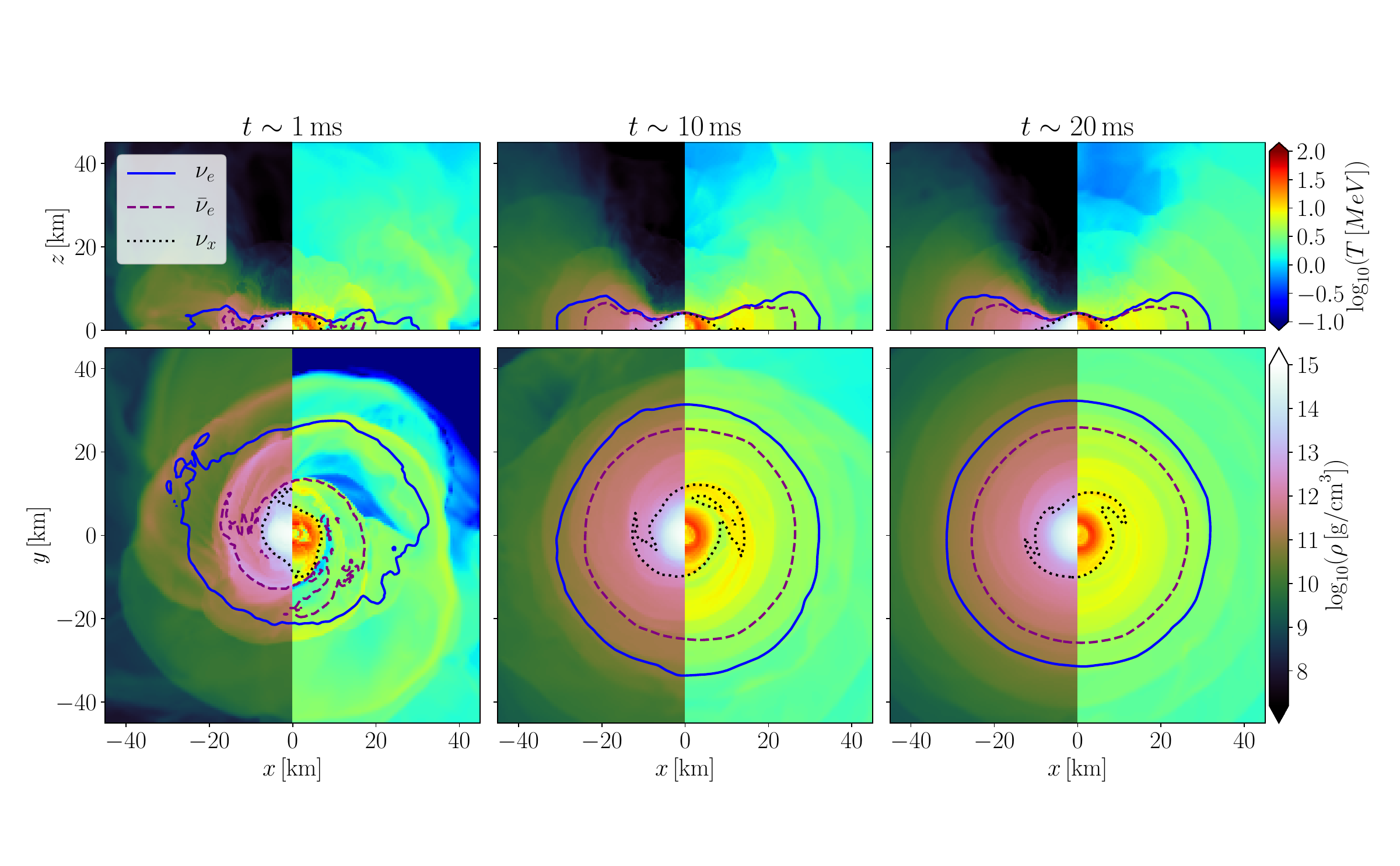}
	\end{center}
	\caption{Same as \Fref{fig:dd2_evol} for the equilibrium neutrino surfaces for the mean neutrino energies at different stages of
          the post-merger of the DD2 model.}
	\label{fig:dd2_evol_eo}
\end{figure*}

\begin{figure*}[!t]
	\begin{center}  
		\includegraphics[width=\textwidth]{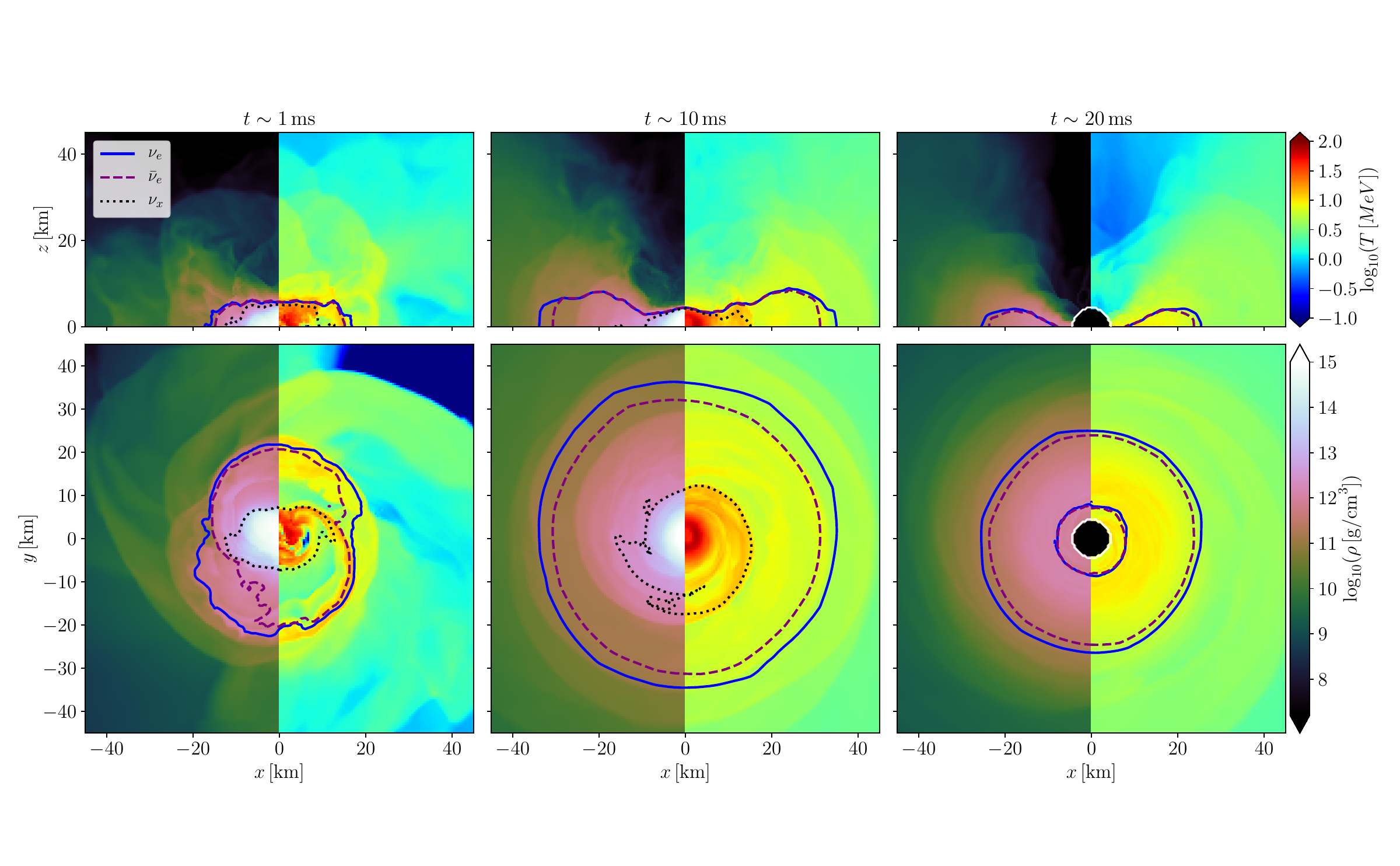}
	\end{center}
	\caption{Same as \Fref{fig:sly4_evol} for the equilibrium neutrino surfaces for the mean neutrino energies at different stages of
          the post-merger of the SLy4 model.}
	\label{fig:sly4_evol_eo}
\end{figure*}


\begin{table*}[!t]
  \caption{Same as \Tref{tab:stau_ev} but for the optical depth $\tau_{eq}$, extracted with \Eref{eq:kappa_eq}.}
    \begin{tabular}{|l|ccc|ccc|}
     \hline
      & & DD2 & & & SLy4 & \\
     \hline
      & $t\sim1$ ms & $t\sim10$ ms & $t\sim20$ ms &
        $t\sim1$ ms & $t\sim10$ ms & $t\sim20$ ms \\
     \hline
     $\log_{10}(\rho^{\nu_e})$ & 
        $11.02\pm0.39$ & $11.02\pm0.25$ & $11.04\pm0.27$ &
        $11.45\pm0.23$ & $11.04\pm0.23$ & $11.44\pm0.25$ \\
     $\log_{10}(T^{\nu_e})$ & 
        $0.48\pm0.17$ & $0.60\pm0.10$ & $0.60\pm0.08$ &
        $0.79\pm0.22$ & $0.76\pm0.07$ & $0.78\pm0.07$ \\
     $Y_e^{\nu_e}$ &
        $0.13\pm0.05$ & $0.14\pm0.02$ & $0.14\pm0.01$ &
        $0.18\pm0.09$ & $0.25\pm0.02$ & $0.18\pm0.01$ \\
     \hline
     $\log_{10}(\rho^{\bar{\nu}_e})$ & 
        $11.79\pm0.31$ & $11.30\pm0.20$ & $11.34\pm0.22$ &
        $11.63\pm0.26$ & $11.14\pm0.21$ & $11.53\pm0.21$ \\
     $\log_{10}(T^{\bar{\nu}_e})$ & 
        $0.61\pm0.17$ & $0.70\pm0.08$ & $0.69\pm0.07$ &
        $0.79\pm0.24$ & $0.79\pm0.06$ & $0.81\pm0.06$ \\
     $Y_e^{\bar{\nu}_e}$ &
        $0.11\pm0.05$ & $0.16\pm0.02$ & $0.16\pm0.01$ &
        $0.16\pm0.08$ & $0.25\pm0.02$ & $0.18\pm0.01$ \\
	 \hline
     $\log_{10}(\rho^{\nu_x})$ & 
        $12.97\pm0.25$ & $12.40\pm0.38$ & $12.74\pm0.22$ &
        $12.54\pm0.27$ & $12.45\pm0.19$ & $-$ \\
     $\log_{10}(T^{\nu_x})$ & 
        $0.96\pm0.15$ & $0.95\pm0.06$ & $0.95\pm0.05$ &
        $1.01\pm0.22$ & $0.98\pm0.10$ & $-$ \\
     $Y_e^{\nu_x}$ &
        $0.08\pm0.03$ & $0.10\pm0.02$ & $0.08\pm0.01$ &
        $0.12\pm0.07$ & $0.11\pm0.03$ & $-$ \\
    \hline
    \end{tabular}
    \centering
  \label{tab:etau_ev}
\end{table*}

\subsubsection{Equilibrium surfaces}

The surfaces where the freeze-out from weak and thermal equilibrium occurs for mean energy neutrinos ({\it i.e.} $\tau_{\rm eq}(\langle E_{\nu} \rangle ) = 2/3$) are shown in \Fref{fig:dd2_evol_eo} and \Fref{fig:sly4_evol_eo}. The difference in the location between the scattering and equilibrium surfaces defines the extension of the diffusion atmosphere, {\it i.e.} the region of the disk where neutrinos still diffuse due to scattering processes up to their diffusion surfaces, but out of local thermodynamics and weak equilibrium. Regardless of the EOS, we find an order inversion between the neutrino species with respect to the scattering surfaces: heavy flavor neutrinos now decouple first, while the electron neutrinos remain longer in equilibrium. This is due to the different behavior of absorption processes involving different neutrino species. In the case of $\nu_e$'s, the dominant equilibration process is the absorption on free neutrons. Due to the high neutron abundance inside the disk, this process effectively provides neutrino opacity almost up to the point where neutrino free streaming occurs. As a consequence, the thermodynamics conditions at the $\nu_e$ equilibrium decoupling surfaces are close to the ones obtained at the scattering decoupling surfaces, as reported in \Tref{tab:etau_ev}. On the contrary, the relative paucity of free protons available to absorb $\bar{\nu}_e$ locates their equilibrium decoupling surfaces at slightly larger densities and temperatures than the ones of $\nu_e$'s. A qualitatively different behavior characterizes $\nu_x$'s. In this case, thermal and weak equilibrium is guaranteed by pair processes, {\it e.g.} $\nu+\bar{\nu} \rightarrow e^+ + e^-$.
Due to their strong temperature dependence, the corresponding equilibrium surfaces shrink deeply inside the remnant, at densities $\sim 10^{13}{\rm g~cm^{-3}}$ where temperatures are usually in excess of $8~{\rm MeV}$.
We also notice that after BH formation ({\it i.e.} in the SLy4 model at $20~{\rm ms}$) there is no region where $\tau_{\rm eq,\nu_x} > 2/3$, which means that even for $E_{\nu_x} \geqslant 20\, \mathrm{MeV}$ the combination of scattering and pair processes that enters \Eref{eq:kappa_eq} requires matter with densities around $\sim 10^{13}~\mathrm{g~cm^{-3}}$ to reach high enough opacities for heavy flavor neutrinos to be significantly absorbed. Since after BH formation the density in the disk drops below a few times $10^{12}{\rm g~cm^{-3}}$ there is no region in the disk where equilibration can occur for heavy neutrinos with average energy.

We have also inspected the volumetric distribution of matter across the equilibrium neutrino surfaces in the $(\rho,T)$ plane and for the DD2 model. In the case of $\nu_e$'s the distributions are practically identical to the one extracted for the
scattering surfaces. For $\bar{\nu}_e$'s, the correlation between $\rho$ and $T$ is even tighter than for the scattering surfaces, even if the surfaces are located at slightly larger densities. Qualitative difference are visible in the case of $\nu_x$'s, where the more relevant role of the matter temperature in the decoupling conditions emerges from the distributions, in particular at late times.

Finally, the values of the temperature at the equilibrium decoupling surfaces allow us to test our initial assumptions about the values of the mean neutrino energies. In fact, neutrinos decoupling first are emitted with higher energies, while neutrinos decoupling at lower densities and temperatures will have a lower average energy. More precisely, the expected relation between the mean energies and the decoupling temperature, $\langle \epsilon_{\nu} \rangle \sim F_2(0)\, T$ (where $F_2(0) \approx 3.15$ is the Fermi integral of order 2 evaluated for 0 degeneracy parameter), is approximately verified.
The agreement is better for the DD2 model, while the softer SLy4 EOS produces a hotter remnant and thus possibly harder neutrino spectra emerging from the equilibrium decoupling surfaces.

\subsection{Neutrino energy dependence of the decoupling conditions}
\label{subs:en_dep}

We now turn to the study of dependency of the neutrino surfaces (and therefore of their thermodynamics properties) on the neutrino energy.  To do that we have extracted the diffusion and the equilibrium optical depths at 20~ms after merger for 8 different energy bins between 3 and 80.33 MeV. We have analyzed both models and for each of them we have obtained different sets of thermodynamics conditions for each neutrino species and optical depth kind. 

\subsubsection{Diffusion surfaces}

\begin{figure*}[!t]
	\begin{center}  
		\includegraphics[width=\textwidth]{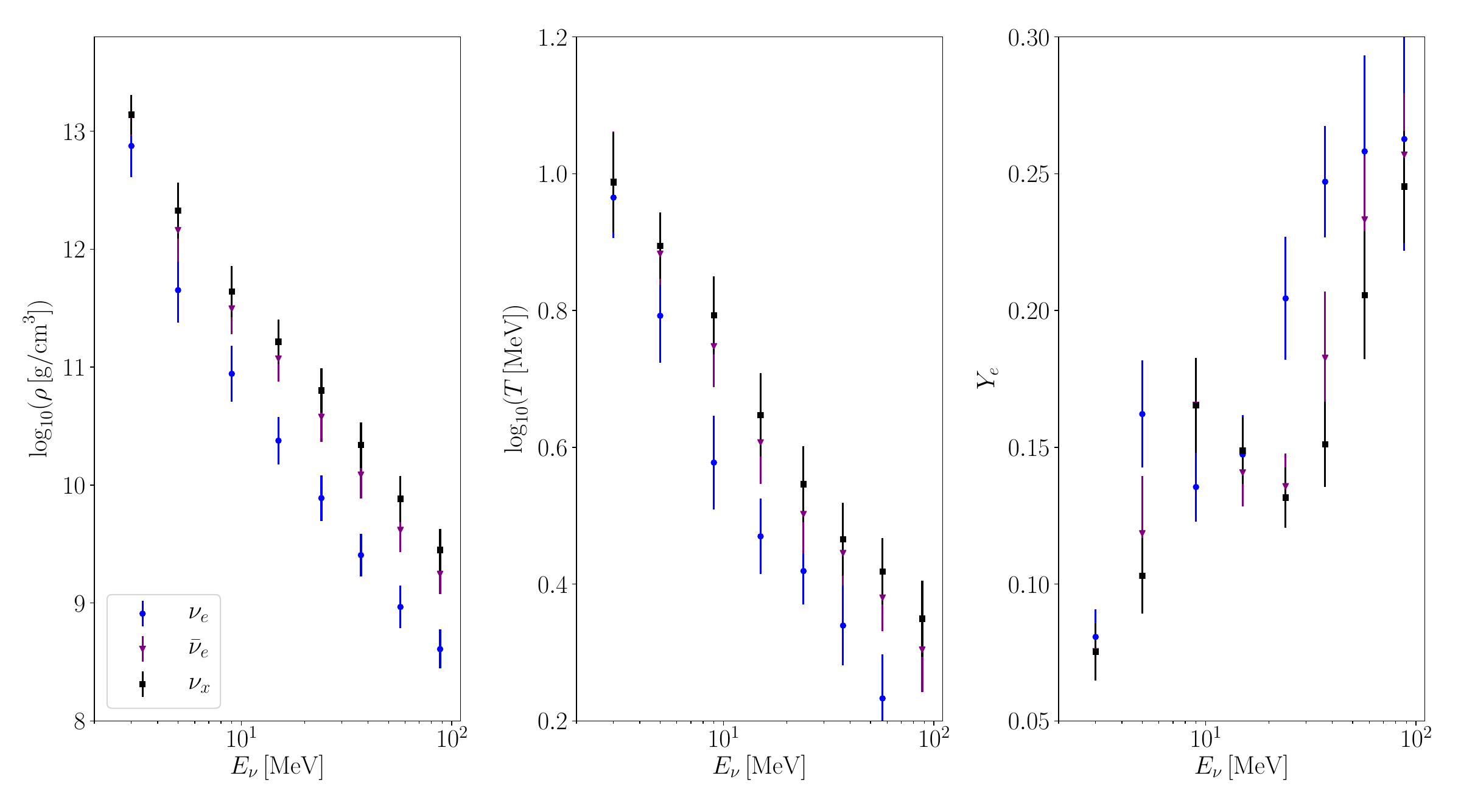}
	\end{center}
	\caption{Rest mass density (left), temperature (center) and electron fraction (right) around the diffusion surfaces for neutrinos with different energies $E_\nu$ for the DD2 model at $20\, \mathrm{ms}$ after merger. Different neutrino species are represented by different markers and colors.}
	\label{fig:opt_en_sc_dd2}
\end{figure*}

\begin{figure*}[!t]
	\begin{center}  
		\includegraphics[width=\textwidth]{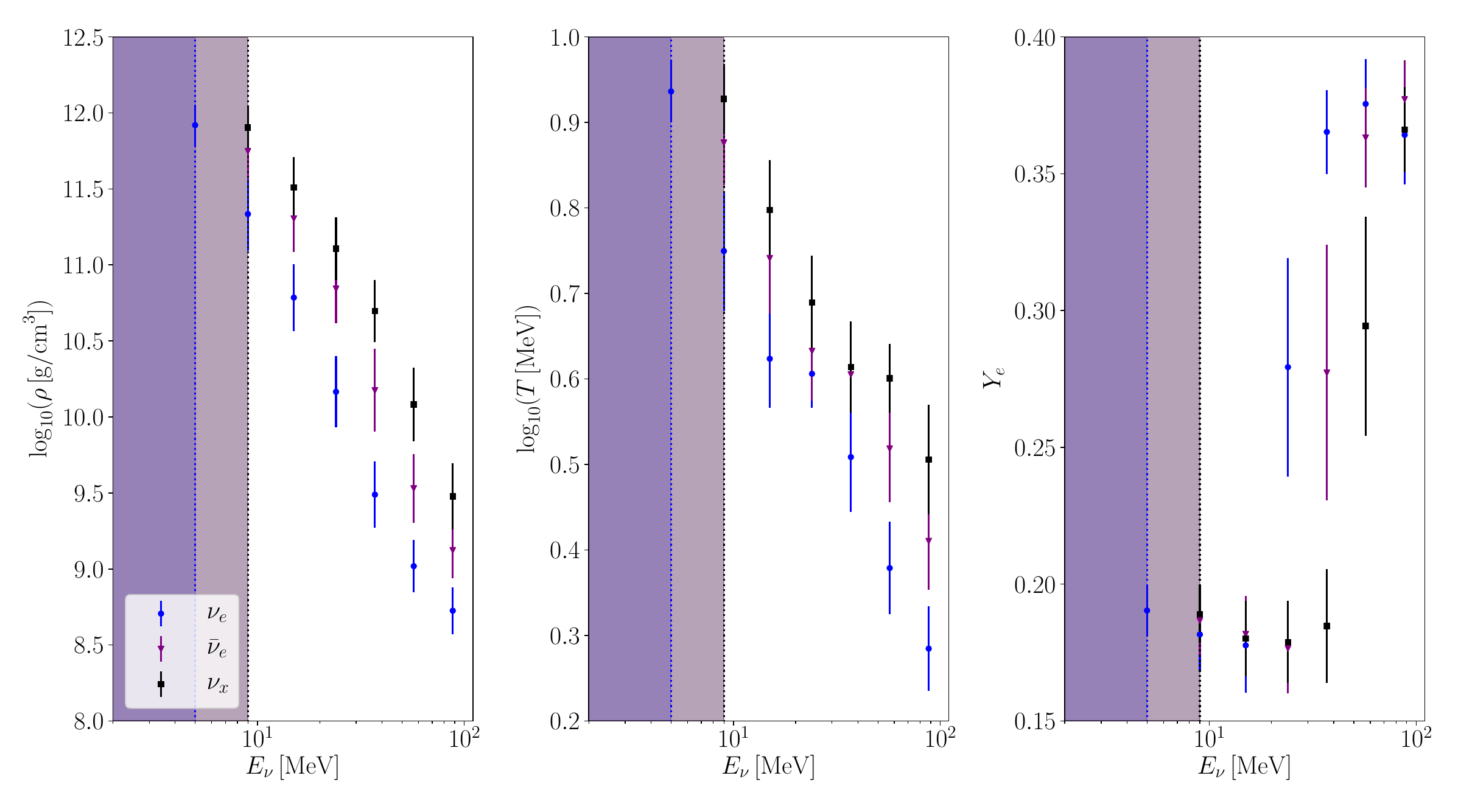}
	\end{center}
	\caption{Same as \Fref{fig:opt_en_sc_dd2} around the diffusion neutrino surfaces for the SLy4 model. In this case, the colored bands represent the neutrino energies for which no optically thick region is found.}
	\label{fig:opt_en_sc_sly}
\end{figure*}

In Figures~\ref{fig:opt_en_sc_dd2}-\ref{fig:opt_en_sc_sly} we show mean densities, temperatures and electron fractions (with their respective standard deviation) around the diffusion surface as function of the neutrino energy, for the DD2 and SLy4 model, respectively.
Due to the $\sim n_{\rm target} E_{\nu}^2$ dependence of the opacity for nucleon scattering and absorption on free baryons, lower energy neutrinos decouple at higher rest mass densities. Since inside the remnant matter temperature increases for increasing density (at least, up to nuclear saturation density, \cite{Perego:2019adq}), the decoupling temperature also decreases monotonically with $E_{\nu}$. Matter deep inside the remnant is closer to weak equilibrium $Y_e$, while the electron fraction inside the disk has been increased by neutrino emission and absorption processes up to 0.30-0.35 in the outer regions of the disk. Thus, $Y_e$ at the decoupling surface for high energy neutrinos is significantly larger than the one of low energy neutrinos.
The two orders of magnitude explored in $E_{\nu}$ translate into four orders of magnitude in decoupling density. Deviations from this leading dependence are visible in the low energy tail, due to final state Pauli-blocking and to second order effects in the interaction cross-section, {\it e.g.} finite electron mass correction. Moreover, for a fixed neutrino energy, the more significant contribution to the total opacity provided by absorption processes, and in particular by the more abundant neutrons, results in a larger inverse mean free path for $\nu_e$'s than for $\bar{\nu}_e$'s and $\nu_x$'s and, consequently, larger decoupling densities and temperatures.

While in the DD2 model (comprising a massive NS in the center) it is always possible to find a scattering decoupling region for all investigated neutrino energies, this is not always the case in the SLy4 model. After the BH has formed, the high density part of the remnant falls inside the AH. In \Fref{fig:opt_en_sc_sly} we do not find a diffusion surface for either $\bar{\nu}_e$ or $\nu_x$ with energies of $5\, \mathrm{MeV}$ or lower. For $\nu_e$ we are able to find a small region that satisfies the condition $0.5<\tau_{\rm diff}<0.85$, but $\sim 90\%$ of it comes from matter at $\tau<2/3$ located in a rather small volume. 


\subsubsection{Equilibrium surfaces}

\begin{figure*}[!t]
	\begin{center}  
		\includegraphics[width=\textwidth]{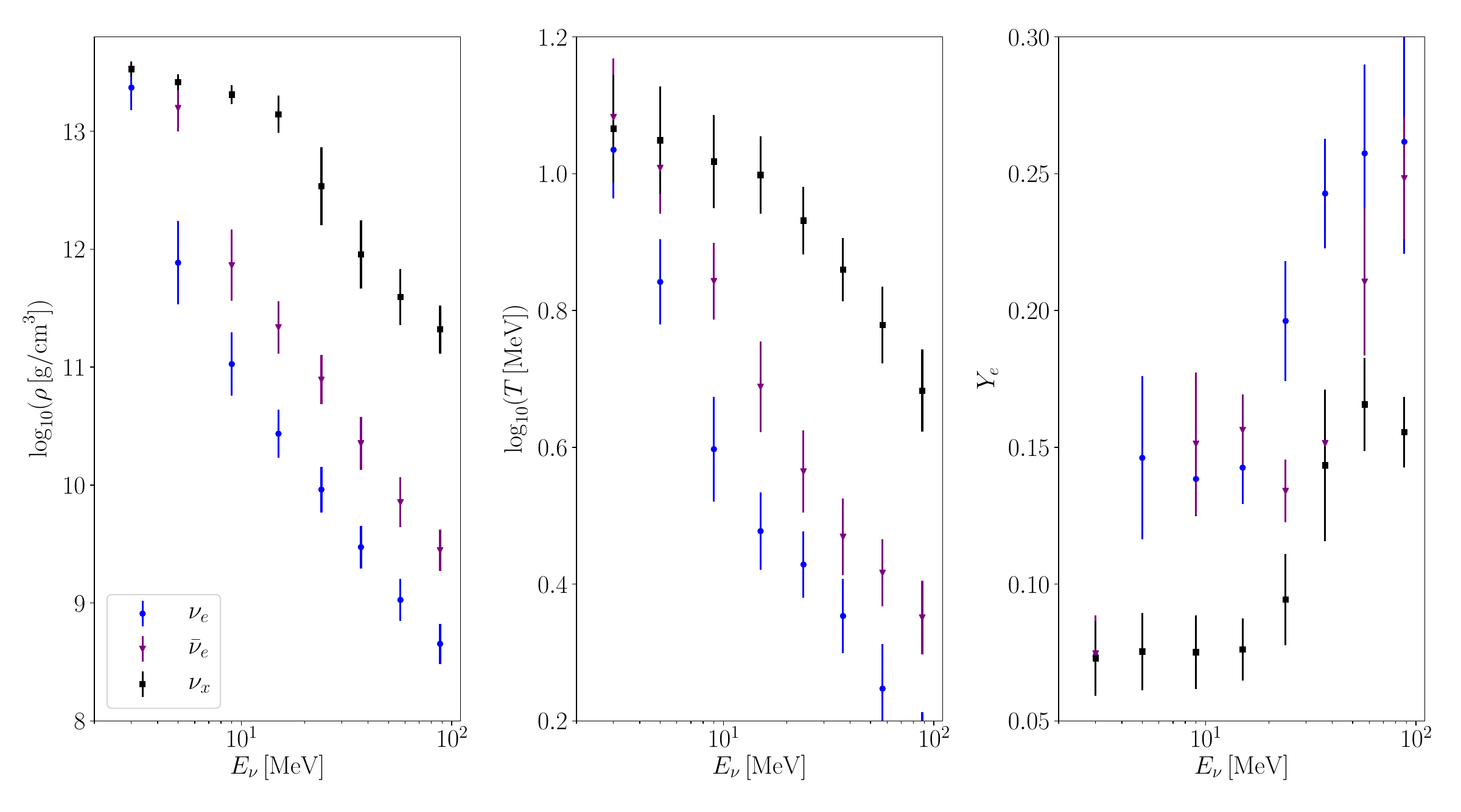}
	\end{center}
	\caption{Rest mass density (left), temperature (center) and electron fraction (right) around the equilibrium surfaces for neutrinos with different energies $E_\nu$ for the DD2 model at $20\, \mathrm{ms}$ after merger. Different neutrino species are represented by different markers and colors.}
	\label{fig:opt_en_eq_dd2}
\end{figure*}

\begin{figure*}[!t]
	\begin{center}  
		\includegraphics[width=\textwidth]{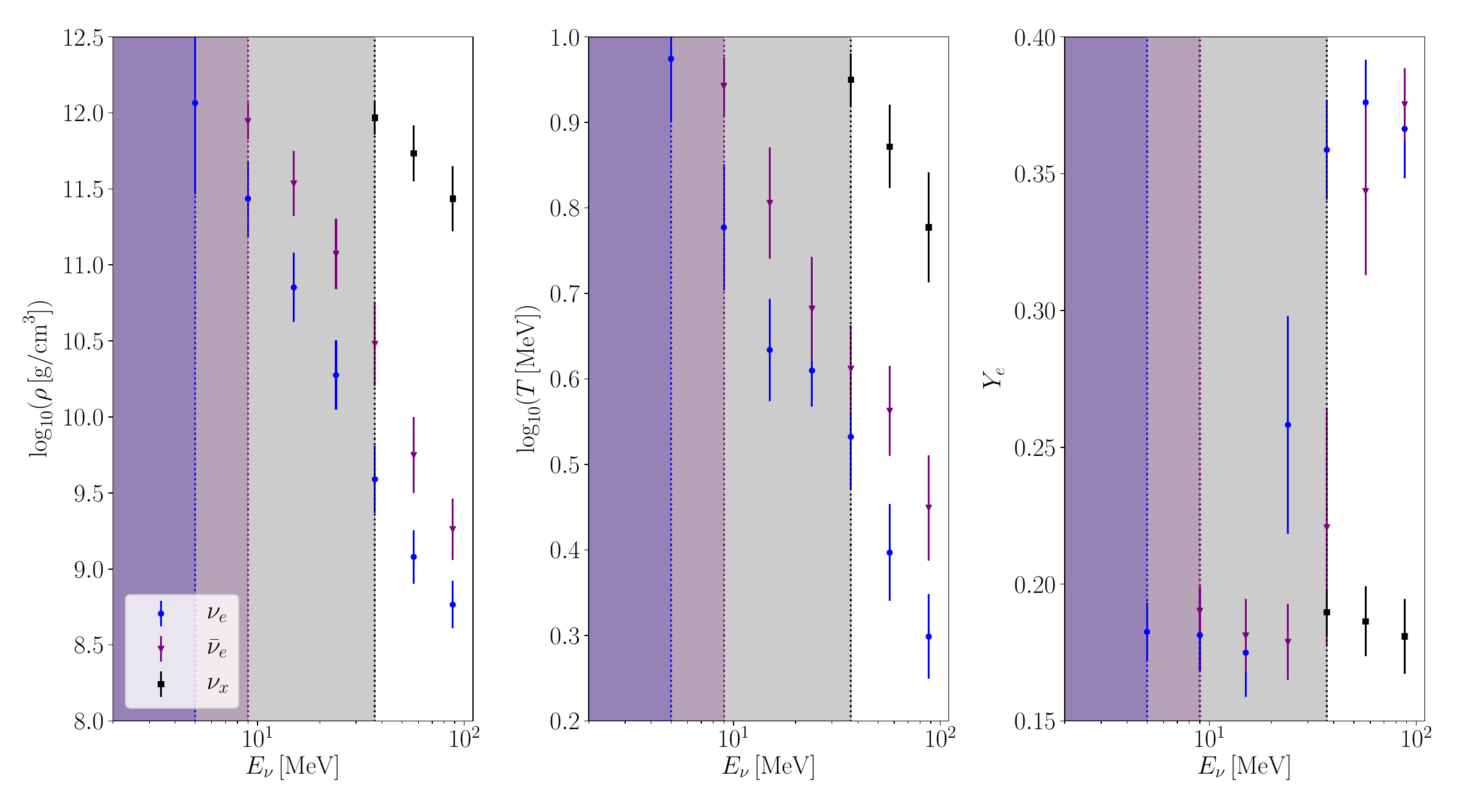}
	\end{center}
	\caption{Same as \Fref{fig:opt_en_eq_dd2} around the equilibrium neutrino surfaces for the SLy4 model. In this case, the colored bands represent the neutrino energies for which no optically thick region is found.}
	\label{fig:opt_en_eq_sly}
\end{figure*}

In \Fref{fig:opt_en_eq_dd2}-\ref{fig:opt_en_eq_sly} we show the result obtained for calculations at different energies of the equilibrium surfaces. Many features observed for the scattering surfaces can be observed also for the equilibrium ones, including the significant dependence on the matter rest mass density.
Since $k_{\rm eq} \leq k_{\rm diff}$, equilibrium decoupling happens always at larger densities and temperatures than the diffusion decoupling, for every neutrino energy. Differences in the decoupling densities are of the order of a few, while they are of 10-15\% in the decoupling temperatures.
Significant deviations from the $\nu_e$ and $\bar{\nu}_e$ behavior can be observed in the case of $\nu_x$ equilibrium surfaces, for which the relevant absorption processes (neutrino pair annihilation and inverse nucleon-nucleon bremsstrahlung) move the decoupling at systematically larger temperatures and densities. In the massive NS case, the decoupling conditions of the three neutrino species become similar for low energy neutrinos. This is an indication that at those conditions pair absorption processes are the most relevant reactions also for $\nu_e$ and $\bar{\nu}_e$. However, due to the low neutrino degeneracy, to the more abundant neutrons and to the larger abundance of electron antineutrinos for $\rho \gtrsim 10^{13}{\rm g~cm^{-3}}$, $T \gtrsim 10~{\rm MeV}$ and $Y_e \sim 0.07$, electron neutrinos decouple again at slightly lower rest mass densities and temperature than $\bar{\nu}_e$ and $\nu_x$.
In the BH remnant case, the lack of equilibrium decoupling surfaces for low energy neutrinos is more severe. This holds especially true for $\nu_x$, whose equilibrium reactions require very high temperatures and rest mass densities to be able to keep low-energy neutrinos in thermal equilibrium. This is visible in the more extended shaded areas in \Fref{fig:opt_en_eq_sly}.

\subsection{Energy-integrated approach}
\label{subs:gray}

\begin{figure*}[!t]
	\begin{center}  
		\includegraphics[width=\textwidth]{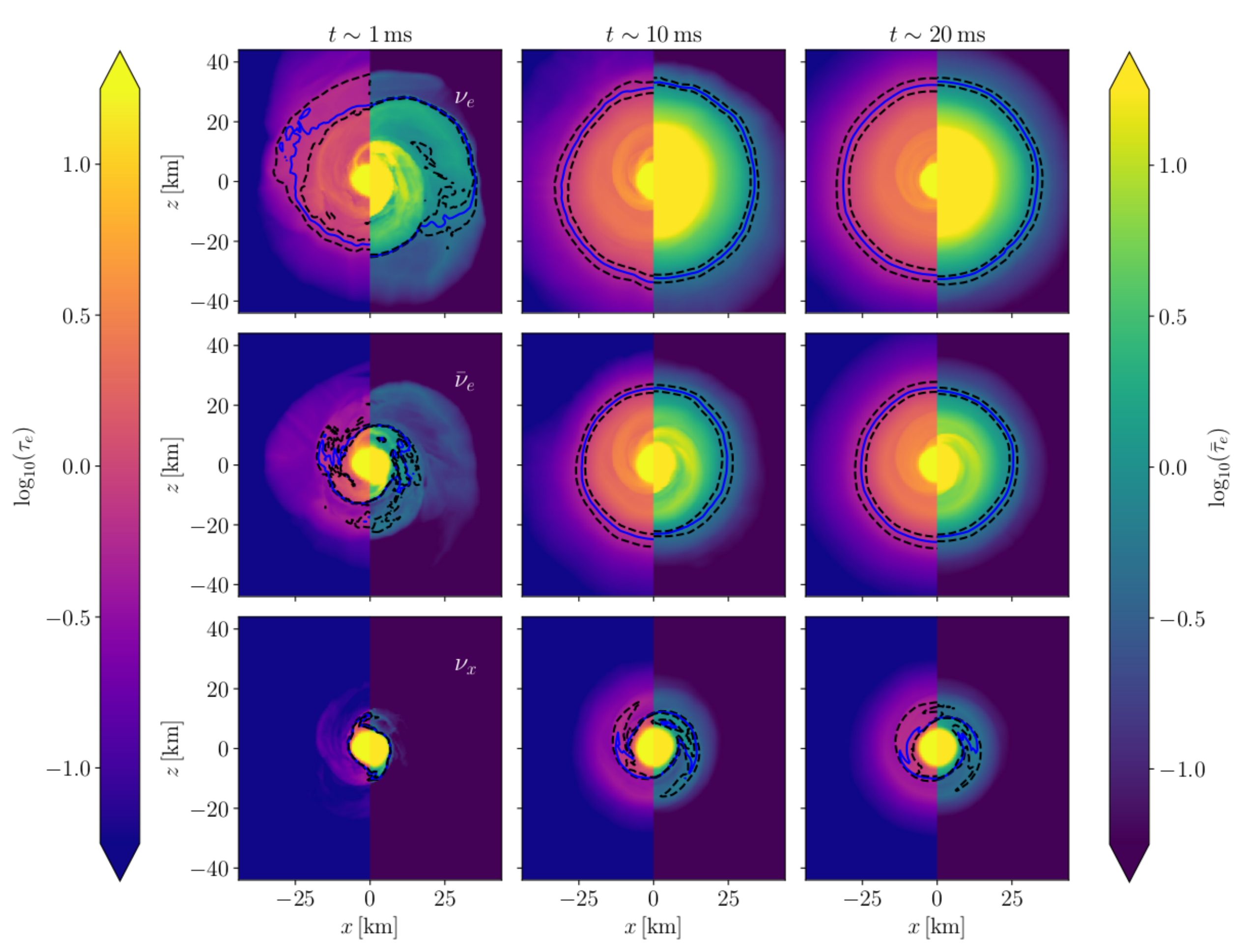}
	\end{center}
	\caption{Equilibrium optical depth for the DD2 model at $20\, \mathrm{ms}$ after merger, evaluated with energy-dependent (left side) and energy-integrated (right side) opacities for each neutrino species ($\nu_e$, top; $\bar{\nu}_e$, middle; $\nu_x$, bottom) on the equatorial plane. The energy dependent opacities are evaluated at the mean neutrino energies at infinity. The solid blue contours represent the surface where $\tau_{\rm eq} \, (\bar{\tau}_{\rm eq})\, = 2/3$, while the black dashed lines are the boundaries of the regions where $0.5<\tau_{\rm eq}\, (\bar{\tau}_{\rm eq})\, <0.85$.}
	\label{fig:dd2_fVe_en}
\end{figure*}

\begin{figure*}[!t]
	\begin{center}  
		\includegraphics[width=\textwidth]{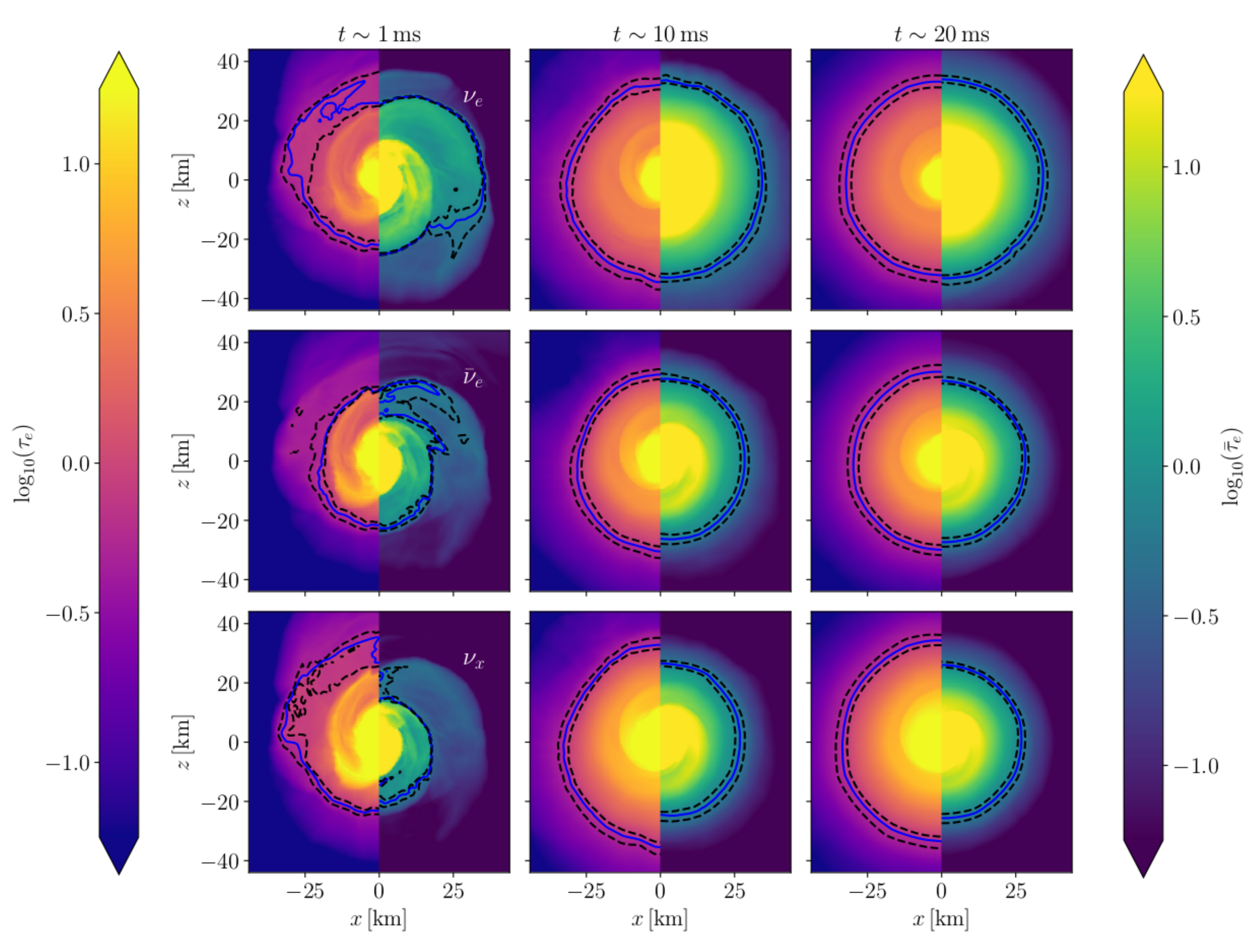}
	\end{center}
	\caption{Same as \Fref{fig:dd2_fVe_en} for the diffusion optical depths $\tau_{\rm diff}$ (left side) and $\bar{\tau}_{\rm diff}$ (right side).}
	\label{fig:dd2_fVe_sc}
\end{figure*}

Finally, to test the robustness and the coherence of the different approaches used in the calculations of the neutrino opacities, we perform the computation of the energy-integrated optical depth $\bar{\tau}$ for both opacity prescriptions, \Eref{eq:kappa_diff} and \Eref{eq:kappa_eq},
and we compare them with the energy-dependent calculations evaluated for the mean neutrino energies at infinity,
$\tau(\langle E_{\nu} \rangle)$. 

The energy integrated approach assumes the neutrino radiation to be in equilibrium with matter. Thus we expect it to be more coherent with the $\tau_{\rm eq}$ case. Indeed, in \Fref{fig:dd2_fVe_en} we show $\tau_{\rm eq}(\langle E_{\nu} \rangle)$ alongside $\bar{\tau}_{\rm eq}$. For concreteness we consider the DD2 model at each of the three snapshots. The location of the neutrino surfaces is remarkably close among the two approaches, in particular after the initial transient phase. We additionally note that the actual values of $\tau$ inside the neutrino surfaces can differ significantly among the two cases (and in particular for the $\nu_e$ case). This is a consequence of the constant neutrino energy used in the energy-dependent approach, while in the energy integrated one the relation between the neutrino energy and the local matter temperature and chemical potentials is more correctly taken into account.
When considering the diffusion optical depth, the hypothesis of equilibrium between matter and radiation is not always verifies.
Consequently in \Fref{fig:dd2_fVe_sc}, where we show the energy dependent and the energy integrated results for the diffusion optical depth, we observe larger discrepancies between the two approaches.
In particular, for $\nu_e$'s the relevance of the absorption on free neutrons guarantees that the diffusion and equilibrium neutrino surfaces stay close, also in the energy integrated case. However, in
the $\bar{\nu}_e$ case and, even more, in the $\nu_x$ case the assumption of matter-radiation equilibrium even for $\bar{\tau}_{\rm diff}$ decreases the mean energies and lowers the opacity. Thus, the neutrino surfaces are apparently located deeper inside in comparison with the expected locations.

\section{Summary} 
\label{sec:conclusions}

In this paper we have studied the thermodynamical properties of matter in the neutrino decoupling region of binary neutron star mergers remnants. 
We have considered two binary neutron star models, characterized by the same total mass ($2.728~M_{\odot}$) and by two different nuclear EOSs (DD2 and SLy4), both compatible with present nuclear and astrophysical constraints. The stiffer DD2 EOS supports a long-lived remnant, while the softer SLy4 EOS produces a black hole roughly $\sim 10~\mathrm{ms}$ after merger. 
In order to span the dependence on the merger dynamics and on the nature of the remnant we have considered three snapshots at times $1$, $10$, and $20\, \mathrm{ms}$ after merger.
We have investigated two kinds of decoupling surfaces: the surfaces from where neutrinos stream freely (diffusion surfaces) and the ones from where thermal and weak equilibrium freezes out (equilibrium surfaces). To account for the dependence of on the neutrino energy, we have considered both neutrinos mean energies (as previously obtained by numerical simulations including neutrino transport) and a large range of binned neutrino energies.

%

Our major findings are that the rest mass density and the neutrino energies are the most relevant quantities in determining the location of the decoupling surfaces and, consequently, the relevant thermodynamical conditions at decoupling for $\nu_e$ and $\bar{\nu}_e$. This is due to the dominant role of quasi-elastic scattering and charged current absorption reactions on free nucleons. For heavy-flavor neutrinos, weak and thermal equilibrium is guaranteed by pair processes, like inverse nucleon-nucleon bremsstrahlung and $\nu$-$\bar{\nu}$ annihilation. Due to the strong dependence of the target and incident particle densities on matter temperature, the latter becomes a relevant thermodynamics quantity as well.

For each model, the diffusion decoupling surfaces of mean energy neutrinos are characterized by similar conditions among the different neutrino species. This is due to a compensation effect between the larger mean energy associated to heavy flavor neutrinos ($k_{\rm scat} \propto E_{\nu}^2$) and the larger abundance of absorbing neutrons relevant for $\nu_e$.
Typical rest mass densities at the corresponding diffusion surfaces are $\sim 10^{11}{\rm g~cm^{-3}}$, with slightly smaller values obtained for $\nu_x$. Temperatures can significantly differ based on the properties of the underlying nuclear EOS. Softer EOSs, like SLy4, produce hotter remnant than stiffer ones, like DD2. In the former case, the transition from diffusion to free streaming regimes for mean energy neutrinos happens between $4$ and $5\, \mathrm{MeV}$, while in the latter between $3$ and $4\, \mathrm{MeV}$. 

Qualitative differences are observed when moving from diffusion to equilibrium surfaces for mean energy neutrinos. On the one hand, due to the neutron richness of the plasma, thermodynamics conditions at the equilibrium and diffusion surfaces are very close for $\nu_e$'s. On the other hand, the relative paucity of absorbing protons moves the equilibrium surfaces of $\bar{\nu}_e$'s deeper inside the remnant, at rest mass densities of a few times $10^{11}{\rm g~cm^{-3}}$. Consequently, the $\bar{\nu}_e$ equilibrium decoupling temperature increases to $5\, \mathrm{MeV}$ in the DD2 case and to $6\, \mathrm{MeV}$ in the SLy4 case. For heavy flavor neutrinos, matter temperature becomes the most relevant quantity and the equilibrium decoupling happens around $10~{\rm MeV}$. Inside the remnant profile this corresponds to a rest mass density in excess of $10^{12}{\rm g~cm^{-3}}$.
Our results for the equilibrium decoupling temperatures broadly satisfy the expected blackbody emission relation, {\it i.e.}
$\langle E_{\nu} \rangle \sim 3.15 T$, with a closer agreement in the case of the stiff EOS model DD2. Deviations from it largely depend on the neutrino emission coming from the volume outside the neutrino surfaces (which is in fact more relevant for hotter remnants). We also notice that in the case of a softer EOS differences between $\nu_e$ and $\bar{\nu}_e$ equilibrium decoupling temperatures reduces, as also observed in \cite{Sekiguchi:2016bjd}.

We have generalized our analysis by considering neutrinos with a broad range of energies, between $3$ and $88\, \mathrm{MeV}$, covering the most relevant part of the emitted spectrum, but irrespective of their relevance. Due to the strong dependence on $E_{\nu}$ of all cross sections, such a large energy range translates in broad ranges of decoupling radii and thermodynamics conditions. In particular, the rest mass density at the diffusion decoupling surface increases between a few times $10^8{\rm g~cm^{-3}}$ for hard neutrinos to  $10^{13} {\rm g~cm^{-3}}$ for soft ones, with a systematic increase going from $\nu_e$ to $\bar{\nu}_e$, and finally to $\nu_x$. The corresponding decoupling temperatures cover one order of magnitude, moving from $\sim2$ to $\sim10\, \mathrm{MeV}$. Similar trends are observed also for electron (anti)neutrinos in the case of the equilibrium surfaces, just for densities and temperatures larger by a factor of a few and $10-15\%$, respectively. A qualitatively different behavior is observed for the equilibrium decoupling of heavy flavor neutrinos, for which temperatures in the range $5$-$12\, \mathrm{MeV}$ are observed, usually increasing the equilibrium decoupling density by one order of magnitude. Important differences can be observed by comparing the DD2 and the SLy4 models. In addition to the larger decoupling temperatures and electron fractions that characterize the SLy4 model (which are a consequence of the hotter and less neutron-rich remnant emerging from softer EOS models), the presence of a massive NS or of a BH in the center largely affects the possibility for the low energy part of the spectrum to thermalize. Indeed, after BH formation, low energy neutrinos do not have decoupling surfaces. This is a consequence of the reduction of the maximum rest mass density and temperature inside the remnant once an AH has formed. The effect is more severe for $\bar{\nu}_e$ ($E_{\nu} \lesssim 10\, \mathrm{MeV}$) and even more for ${\nu}_x$ ($E_{\nu} \lesssim 40\, \mathrm{MeV}$) than for ${\nu}_e$ ($E_{\nu} \lesssim 5\, \mathrm{MeV}$).

As a final check, we have computed the optical depths (and the corresponding decoupling surfaces) using an energy integrated approach that assumes everywhere equilibrium between the neutrino field and the matter. We have verified that the equilibrium neutrino surfaces obtained with the energy dependent approach (evaluated for the mean neutrino energies) and with the energy-integrated one lie very close, while relevant differences are observed in the case of the diffusion surfaces.

All the results we have reported are robust with respect to the optical depth calculation algorithm and to the grid resolution. However, all our simulations emply an approximate neutrino transport. While the present leakage+M0 treatment is expected to catch the dominant cooling features of neutrino emission, the lack of an explicit modelling of trapped neutrinos and uncertainties in the challenging radiation transport can affect more significantly the electron fraction inside the remnant. Thus, we expect the values of the density and temperature at the decoupling surfaces to be more robust than the ones of the electron fraction.

Our study represents one of the first systematic investigations of neutrino opacities in binary neutron star mergers and of the thermodynamics conditions occurring where neutrino decouple from matter. Our results have revealed how the large variety of thermodynamics conditions inside the remnant, combined with
its disk geometry, translates in broad ranges of decoupling conditions in all the relevant thermodynamics quantities. These can influence the properties of the emitted spectrum and ultimately of many multimessenger observables through the evolution of the remnant and of the ejecta. 
Matter conditions predicted for BNS mergers share many similarities with the ones experienced by matter in the gravitational collapse of the iron core of massive stars. Indeed, also in CCSNe neutrinos of all flavors are copiously produced and become trapped at large enough densities on time scales much larger than the dynamical (and even the explosion) time scale. Due to the high sensitivity of CCSN explosions on weak reactions, neutrino-matter interactions in CCSNe have been deeply investigated
in many studies, {\it e.g.} \cite{Roberts.etal:2012,Lentz:2012xc,Abdikamalov:2014oba,Melson:2015,Fischer:2016,OConnor:2018sti}. In particular, the densities and temperatures where neutrinos stream freely or decouple from matter are similar to the ones we found in BNS mergers. During the collapse before core bounce, $Y_e \gtrsim 0.3$ and the maximum core temperature (and thus the neutrino mean energies) stay low, $\lesssim$ a few MeV. Under these conditions, decoupling occurs at higher densities (e.g., several times $10^{11}{\rm g~cm^{-3}}$ for mean energy $\nu_e$'s), lower temperatures and higher $Y_{e}$. After core bounce, during the accretion phase, matter and neutrino temperatures increase at the surface of the forming PNS. As a consequence, the decoupling densities and temperatures of CCSNe become more similar to the ones expected in BNS mergers. In the case of very massive zero-age main sequence stars ($\gtrsim 25 M_{\odot}$), even the maximum temperatures are comparable to the one observed in BNS mergers. During the subsequent PNS cooling phase, decoupling temperatures progressively decrease and the neutrino surfaces sink again towards larger densities.
However, the most noteworthy difference between neutrino surfaces in CCSNe and BNS mergers resides in the surface geometry: while in CCSNe they are all spheroidal, in BNS merger remnants containing a massive NS they are close to spherical geometry for low energy neutrinos and increase to a disk-like shape inside the disk for larger neutrino energies.

In light of our results, the accurate modelling of neutrinos and of their emitted spectrum in BNS mergers requires the consistent inclusion of neutrino-matter interactions for a broad range of nuclear conditions, characterized by more than four orders of magnitude in rest mass densities, one order of magnitude in temperatures, and a factor of a few in electron fraction.
This would require the usage in merger simulations of more consistent and accurate weak reaction rates, {\it e.g.} \cite{Bacca:2011qd,Rrapaj:2014yba,Roberts:2016mwj,Guo:2019cvs} .
This represents a great challenge in nuclear and neutrino physics for the years to come.

\paragraph*{Acknowledgments}

The authors thank Bernd Br\"ugmann and Domenico Logoteta for useful discussions.
The authors thank the organizers and participants 
of the INT Program INT-18-72R ``First Multi-Messenger Observation of a Neutron Star Merger
and its Implications for Nuclear Physics INT workshop'' held at
(Seattle, March 2018), 
of the ExtreMe Matter Institute's rapid task force meeting at
GSI/FAIR (Darmstadt, June 2018),
of the GWEOS workshop (Pisa, February 2019),  
for stimulating discussions.
SB acknowledges support by the EU H2020 under ERC Starting Grant, 
no.~BinGraSp-714626. 
DR acknowledges support from a Frank and Peggy Taplin Membership at the
Institute for Advanced Study and the
Max-Planck/Princeton Center (MPPC) for Plasma Physics (NSF PHY-1804048).
Computations were performed 
on the supercomputer SuperMUC at the LRZ Munich (Gauss project
pn56zo), 
on supercomputer Marconi at CINECA (ISCRA-B project number HP10B2PL6K
and HP10BMHFQQ); on the supercomputers Bridges, Comet, and Stampede (NSF XSEDE allocation TG-PHY160025);
on NSF/NCSA Blue Waters (NSF AWD-1811236).
AE, AP, and BG acknowledge computational support also by the INFN initiative TEONGRAV. 
This research used resources of the National Energy Research
Scientific Computing Center, a DOE Office of Science User Facility supported by the Office of Science of the U.S. Department of Energy under Contract No. DE-AC02-05CH11231.

%
% ...
%

\appendix

\section{Scheme convergence}
\label{app:convergence}
In order to check the robustness of our scheme, we extracted the optical depth over grids with different resolutions (interpolated from the same simulation results). We call standard resolution (SR) the one used throughout the paper ($h'\sim0.74\, \mathrm{km}$); we then picked as our best resolution (HR) a grid with $h'\sim0.49\, \mathrm{km}$ and as our worst resolution (LR) a grid with twice the spacing of our SR ($h'\sim1.48\, \mathrm{km}$). In order to make a more informed convergence study, we also considered a medium resolution (MR) grid, with $h'\sim1.11\, \mathrm{km}$. 

\begin{table*}[!t]
  \caption{In this table we present the mean values of density $\rho$, temperature $T$ and electron fraction $Y_e$ in the regions with opacity $0.5<\tau<0.85$ using \Eref{eq:kappa_diff} for four different resolutions. The considered model is the DD2 at $20\, \mathrm{ms}$ after merger. Densities are given as $\log_{10}(\rho\, [\mathrm{g/cm}^3])$ and temperatures are expressed in MeV. For each quantity we also give its variance in the distribution.}
    \begin{tabular}{|l|cccc|}
     \hline
      & LR $(1.48\, \mathrm{km})$ & MR $(1.11\, \mathrm{km})$ &
       SR $(0.74\, \mathrm{km})$ & HR $(0.49\, \mathrm{km})$ \\
     \hline
     $\log_{10}(\rho^{\nu_e})$ & 
        $11.13\pm0.27$ & $11.16\pm0.28$ & $11.18\pm0.29$ &
        $11.24\pm0.33$ \\
     $\log_{10}(T^{\nu_e})$ & 
        $0.62\pm0.07$ & $0.63\pm0.07$ & $0.63\pm0.07$ &
        $0.64\pm0.08$ \\
     $Y_e^{\nu_e}$ &
        $0.13\pm0.01$ & $0.13\pm0.01$ & $0.13\pm0.01$ &
        $0.13\pm0.01$ \\
     \hline
     $\log_{10}(\rho^{\bar{\nu}_e})$ & 
        $11.16\pm0.18$ & $11.17\pm0.18$ & $11.21\pm0.21$ &
        $11.22\pm0.21$ \\
     $\log_{10}(T^{\bar{\nu}_e})$ & 
        $0.63\pm0.05$ & $0.63\pm0.06$ & $0.64\pm0.05$ &
        $0.64\pm0.06$ \\
     $Y_e^{\bar{\nu}_e}$ &
        $0.14\pm0.01$ & $0.14\pm0.01$ & $0.15\pm0.01$ &
        $0.14\pm0.01$ \\
	 \hline
     $\log_{10}(\rho^{\nu_x})$ & 
        $10.86\pm0.17$ & $10.90\pm0.19$ & $10.92\pm0.19$ &
        $10.94\pm0.20$ \\
     $\log_{10}(T^{\nu_x})$ & 
        $0.56\pm0.05$ & $0.57\pm0.05$ & $0.57\pm0.05$ &
        $0.58\pm0.05$ \\
     $Y_e^{\nu_x}$ &
        $0.13\pm0.01$ & $0.13\pm0.01$ & $0.13\pm0.01$ &
        $0.13\pm0.01$ \\
    \hline
    \end{tabular}
    \centering
  \label{tab:res_conv}
\end{table*}
In \Tref{tab:res_conv} we give the mean values for the main hydrodynamic quantities in the transition region for the three resolutions considered. We can see how the increase in resolution translates into an increase in the mean values of rest mass density and temperature in the transient region. We expect this to happen due to how optical depth is evaluated. The higher the number of cells in the region where rest mass density varies quickly means that a cell that previously occupied a volume V with density $\rho$ now is split into $8$ cells, increasing the possibility that at least one of them has a density $\rho_1 < \rho$ that is just enough to push the value of $\tau$ for that cell below the threshold. This only works towards decreasing $\tau$, since we always pick the minimum optical depth between all directions considered, and therefore is sufficient that this happens in one direction to decrease $\tau$, while for this mechanism to increase $\tau$ the new grid should have all the neighbouring cells to end up with higher $\rho$. This is unlikely given the distribution of density in the system.

\section{Scheme comparison}
\label{sec:sccomp}

To test the robustness of our results with respect to the optical depth computations, we repeat the calculation of the decoupling thermodynamics conditions for mean energy neutrinos using MODA \cite{Perego:2014qda}. For concreteness, we perform the analysis for the last two simulation snapshots (10 and 20~ms after merger) and we focus on the diffusion optical depth. In \Tref{tab:MODA_stau_ev}
we summarize the thermodynamic decoupling conditions. All the results are compatible with our analysis within uncertainties. 

 \begin{table*}[!t]
  \caption{In this table we present the mean values of density $\rho$, temperature $T$ and electron fraction $Y_e$ in the regions with opacity $0.5\leq\tau_{s}\leq0.85$ using \Eref{eq:kappa_diff} for both models at different stages of the postmerger evolution. In the Sly4 case, at $t\sim20$ ms, the remnant already collapsed to BH. Densities are given as $\log_{10}(\rho\, [\mathrm{g~cm^{-3}}])$ and temperatures as $T\, [\mathrm{MeV}]$. For each quantity we also give its variance in the distribution.}
    \begin{tabular}{|l|cc|cc|}
     \hline
       & \hspace{2cm} DD2  & & \hspace{2cm} SLy4 & \\
     \hline
           & $t\sim10$ ms & $t\sim20$ ms &
           $t\sim10$ ms & $t\sim20$ ms \\
     \hline
     $\log_{10}(\rho^{\nu_e})$ & 
          $11.17\pm0.27$ & $11.21\pm0.26$ &
          $11.04\pm0.25$ & $11.40\pm0.25$ \\
     $\log_{10}(T^{\nu_e})$ & 
	      $0.67\pm0.21$ & $0.66\pm0.18$ &
		  $0.77\pm0.16$ & $0.77\pm0.15$ \\          
     $Y_e^{\nu_e}$ &
         $0.15\pm0.02$ & $0.15\pm0.01$ &
         $0.24\pm0.03$ & $0.18\pm0.01$ \\
     \hline
     $\log_{10}(\rho^{\bar{\nu}_e})$ & 
         $10.91\pm0.24$ & $10.93\pm0.21$ &
         $11.01\pm0.21$ & $11.35\pm0.22$ \\
     $\log_{10}(T^{\bar{\nu}_e})$ & 
         $0.59\pm0.17$ & $0.58\pm0.14$ &
         $0.76\pm0.24$ & $0.76\pm0.13$ \\
     $Y_e^{\bar{\nu}_e}$ &
        $0.14\pm0.02$ & $0.14\pm0.02$ &
        $0.25\pm0.02$ & $0.18\pm0.01$ \\
	 \hline
     $\log_{10}(\rho^{\nu_x})$ & 
         $10.46\pm0.32$ & $10.46\pm0.32$ &
         $10.86\pm0.20$ & $11.15\pm0.25$ \\
     $\log_{10}(T^{\nu_x})$ & 
         $0.50\pm0.19$ & $0.50\pm0.12$ &
         $0.72\pm0.13$ & $0.70\pm0.13$ \\
     $Y_e^{\nu_x}$ &
         $0.14\pm0.03$ & $0.14\pm0.01$ &
         $0.24\pm0.02$ & $0.18\pm0.02$ \\
    \hline
    \end{tabular}
    \centering
  \label{tab:MODA_stau_ev}
\end{table*}

\newpage
\bibliographystyle{epj.bst}
\bibliography{references}

\begin{thebibliography}{88}

\bibitem{Rosswog:2015nja}
S.~Rosswog, Int.J.Mod.Phys. \textbf{D24}, 1530012 (2015), \texttt{1501.02081}

\bibitem{Baiotti:2016qnr}
L.~Baiotti, L.~Rezzolla, Rept. Prog. Phys. \textbf{80}, 096901 (2017),
  \texttt{1607.03540}

\bibitem{Barack:2018yly}
L.~Barack et~al., Class. Quant. Grav. \textbf{36}, 143001 (2019),
  \texttt{1806.05195}

\bibitem{Hotokezaka:2011dh}
K.~Hotokezaka, K.~Kyutoku, H.~Okawa, M.~Shibata, K.~Kiuchi, Phys.Rev.
  \textbf{D83}, 124008 (2011), \texttt{1105.4370}

\bibitem{Bauswein:2013jpa}
A.~Bauswein, T.~Baumgarte, H.T. Janka, Phys.Rev.Lett. \textbf{111}, 131101
  (2013), \texttt{1307.5191}

\bibitem{Hotokezaka:2013iia}
K.~Hotokezaka, K.~Kiuchi, K.~Kyutoku, T.~Muranushi, Y.i. Sekiguchi et~al.,
  Phys.Rev. \textbf{D88}, 044026 (2013), \texttt{1307.5888}

\bibitem{Zappa:2017xba}
F.~Zappa, S.~Bernuzzi, D.~Radice, A.~Perego, T.~Dietrich, Phys. Rev. Lett.
  \textbf{120}, 111101 (2018), \texttt{1712.04267}

\bibitem{Radice:2018pdn}
D.~Radice, A.~Perego, K.~Hotokezaka, S.A. Fromm, S.~Bernuzzi, L.F. Roberts,
  Astrophys. J. \textbf{869}, 130 (2018), \texttt{1809.11161}

\bibitem{Dietrich:2018phi}
T.~Dietrich, D.~Radice, S.~Bernuzzi, F.~Zappa, A.~Perego, B.~Brügmann, S.V.
  Chaurasia, R.~Dudi, W.~Tichy, M.~Ujevic, Class. Quant. Grav. \textbf{35},
  24LT01 (2018), \texttt{1806.01625}

\bibitem{Koppel:2019pys}
S.~Köppel, L.~Bovard, L.~Rezzolla, Astrophys. J. \textbf{872}, L16 (2019),
  \texttt{1901.09977}

\bibitem{Lattimer:1974a}
J.M. {Lattimer}, D.N. {Schramm}, apjl \textbf{192}, L145 (1974)

\bibitem{Symbalisty:1982a}
E.~{Symbalisty}, D.N. {Schramm}, Astrophys. J. Letters \textbf{22}, 143 (1982)

\bibitem{Thielemann:2017acv}
F.K. Thielemann, M.~Eichler, I.V. Panov, B.~Wehmeyer, Ann. Rev. Nucl. Part.
  Sci. \textbf{67}, 253 (2017), \texttt{1710.02142}

\bibitem{Li:1998bw}
L.X. Li, B.~Paczynski, Astrophys.J. \textbf{507}, L59 (1998),
  \texttt{astro-ph/9807272}

\bibitem{Kulkarni:2005jw}
S.R. Kulkarni (2005), \texttt{astro-ph/0510256}

\bibitem{Metzger:2016pju}
B.D. Metzger, Living Rev. Rel. \textbf{20}, 3 (2017), \texttt{1610.09381}

\bibitem{Paczynski:1986px}
B.~Paczynski, Astrophys. J. \textbf{308}, L43 (1986)

\bibitem{Eichler:1989ve}
D.~Eichler, M.~Livio, T.~Piran, D.N. Schramm, Nature \textbf{340}, 126 (1989)

\bibitem{TheLIGOScientific:2017qsa}
B.P. Abbott et~al. (Virgo, LIGO Scientific), Phys. Rev. Lett. \textbf{119},
  161101 (2017), \texttt{1710.05832}

\bibitem{Mooley:2018dlz}
K.P. Mooley, A.T. Deller, O.~Gottlieb, E.~Nakar, G.~Hallinan, S.~Bourke, D.A.
  Frail, A.~Horesh, A.~Corsi, K.~Hotokezaka, Nature \textbf{561}, 355 (2018),
  \texttt{1806.09693}

\bibitem{Ghirlanda.etal:2019}
G.~{Ghirlanda}, O.S. {Salafia}, Z.~{Paragi}, M.~{Giroletti}, J.~{Yang},
  B.~{Marcote}, J.~{Blanchard}, I.~{Agudo}, T.~{An}, M.G. {Bernardini} et~al.,
  Science \textbf{363}, 968 (2019), \texttt{1808.00469}

\bibitem{Monitor:2017mdv}
B.P. Abbott et~al. (Virgo, Fermi-GBM, INTEGRAL, LIGO Scientific), Astrophys. J.
  \textbf{848}, L13 (2017), \texttt{1710.05834}

\bibitem{GBM:2017lvd}
B.P. Abbott et~al. (GROND, SALT Group, OzGrav, DFN, INTEGRAL, Virgo,
  Insight-Hxmt, MAXI Team, Fermi-LAT, J-GEM, RATIR, IceCube, CAASTRO, LWA,
  ePESSTO, GRAWITA, RIMAS, SKA South Africa/MeerKAT, H.E.S.S., 1M2H Team,
  IKI-GW Follow-up, Fermi GBM, Pi of Sky, DWF (Deeper Wider Faster Program),
  Dark Energy Survey, MASTER, AstroSat Cadmium Zinc Telluride Imager Team,
  Swift, Pierre Auger, ASKAP, VINROUGE, JAGWAR, Chandra Team at McGill
  University, TTU-NRAO, GROWTH, AGILE Team, MWA, ATCA, AST3, TOROS, Pan-STARRS,
  NuSTAR, ATLAS Telescopes, BOOTES, CaltechNRAO, LIGO Scientific, High Time
  Resolution Universe Survey, Nordic Optical Telescope, Las Cumbres Observatory
  Group, TZAC Consortium, LOFAR, IPN, DLT40, Texas Tech University, HAWC,
  ANTARES, KU, Dark Energy Camera GW-EM, CALET, Euro VLBI Team, ALMA),
  Astrophys. J. \textbf{848}, L12 (2017), \texttt{1710.05833}

\bibitem{Kasen:2017sxr}
D.~Kasen, B.~Metzger, J.~Barnes, E.~Quataert, E.~Ramirez-Ruiz, Nature  (2017),
  [Nature551,80(2017)], \texttt{1710.05463}

\bibitem{Villar:2017wcc}
V.A. Villar et~al., Astrophys. J. \textbf{851}, L21 (2017), \texttt{1710.11576}

\bibitem{Perego:2017wtu}
A.~Perego, D.~Radice, S.~Bernuzzi, Astrophys. J. \textbf{850}, L37 (2017),
  \texttt{1711.03982}

\bibitem{Wollaeger:2017ahm}
R.T. Wollaeger, O.~Korobkin, C.J. Fontes, S.K. Rosswog, W.P. Even, C.L. Fryer,
  J.~Sollerman, A.L. Hungerford, D.R. van Rossum, A.B. Wollaber (2017),
  \texttt{1705.07084}

\bibitem{Kawaguchi:2018ptg}
K.~Kawaguchi, M.~Shibata, M.~Tanaka, Astrophys. J. \textbf{865}, L21 (2018),
  \texttt{1806.04088}

\bibitem{Miller:2019dpt}
J.M. Miller, B.R. Ryan, J.C. Dolence, A.~Burrows, C.J. Fontes, C.L. Fryer,
  O.~Korobkin, J.~Lippuner, M.R. Mumpower, R.T. Wollaeger (2019),
  \texttt{1905.07477}

\bibitem{Perego:2019adq}
A.~Perego, S.~Bernuzzi, D.~Radice, Eur. Phys. J. \textbf{A55}, 124 (2019),
  \texttt{1903.07898}

\bibitem{Ruffert:1996by}
M.~Ruffert, H.~Janka, K.~Takahashi, G.~Sch{\"a}fer, Astron.Astrophys.
  \textbf{319}, 122 (1997), \texttt{astro-ph/9606181}

\bibitem{Rosswog:2003rv}
S.~Rosswog, M.~Liebendoerfer, Mon.Not.Roy.Astron.Soc. \textbf{342}, 673 (2003),
  \texttt{astro-ph/0302301}

\bibitem{Perego:2017fho}
A.~Perego, H.~Yasin, A.~Arcones, J. Phys. \textbf{G44}, 084007 (2017),
  \texttt{1701.02017}

\bibitem{Dessart:2008zd}
L.~Dessart, C.~Ott, A.~Burrows, S.~Rosswog, E.~Livne, Astrophys.J.
  \textbf{690}, 1681 (2009), \texttt{0806.4380}

\bibitem{Perego:2014fma}
A.~Perego, S.~Rosswog, R.~Cabezon, O.~Korobkin, R.~Kaeppeli et~al.,
  Mon.Not.Roy.Astron.Soc. \textbf{443}, 3134 (2014), \texttt{1405.6730}

\bibitem{Just:2014fka}
O.~Just, A.~Bauswein, R.A. Pulpillo, S.~Goriely, H.T. Janka, Mon. Not. Roy.
  Astron. Soc. \textbf{448}, 541 (2015), \texttt{1406.2687}

\bibitem{Foucart:2015gaa}
F.~Foucart, R.~Haas, M.D. Duez, E.~O'Connor, C.D. Ott, L.~Roberts, L.E. Kidder,
  J.~Lippuner, H.P. Pfeiffer, M.A. Scheel, Phys. Rev. \textbf{D93}, 044019
  (2016), \texttt{1510.06398}

\bibitem{Sekiguchi:2016bjd}
Y.~Sekiguchi, K.~Kiuchi, K.~Kyutoku, M.~Shibata, K.~Taniguchi, Phys. Rev.
  \textbf{D93}, 124046 (2016), \texttt{1603.01918}

\bibitem{Martin:2017dhc}
D.~Martin, A.~Perego, W.~Kastaun, A.~Arcones, Class. Quant. Grav. \textbf{35},
  034001 (2018), \texttt{1710.04900}

\bibitem{Dicus:1972yr}
D.A. Dicus, Phys. Rev. \textbf{D6}, 941 (1972)

\bibitem{Bruenn:1985en}
S.W. Bruenn, Astrophys. J. Suppl. \textbf{58}, 771 (1985)

\bibitem{Formaggio:2013kya}
J.A. Formaggio, G.P. Zeller, Rev. Mod. Phys. \textbf{84}, 1307 (2012),
  \texttt{1305.7513}

\bibitem{Martin:2015hxa}
D.~Martin, A.~Perego, A.~Arcones, F.K. Thielemann, O.~Korobkin, S.~Rosswog,
  Astrophys. J. \textbf{813}, 2 (2015), \texttt{1506.05048}

\bibitem{Metzger:2014ila}
B.D. Metzger, R.~Fern\'{a}ndez, Mon.Not.Roy.Astron.Soc. \textbf{441}, 3444
  (2014), \texttt{1402.4803}

\bibitem{Metzger:2017wot}
B.D. Metzger (2017), \texttt{1710.05931}

\bibitem{Galeazzi:2013mia}
F.~Galeazzi, W.~Kastaun, L.~Rezzolla, J.A. Font, Phys.Rev. \textbf{D88}, 064009
  (2013), \texttt{1306.4953}

\bibitem{Palenzuela:2015dqa}
C.~Palenzuela, S.L. Liebling, D.~Neilsen, L.~Lehner, O.L. Caballero,
  E.~O’Connor, M.~Anderson, Phys. Rev. \textbf{D92}, 044045 (2015),
  \texttt{1505.01607}

\bibitem{Foucart:2016rxm}
F.~Foucart, E.~O'Connor, L.~Roberts, L.E. Kidder, H.P. Pfeiffer, M.A. Scheel,
  Phys. Rev. \textbf{D94}, 123016 (2016), \texttt{1607.07450}

\bibitem{Perego:2015agy}
A.~Perego, R.~Cabezon, R.~Kaeppeli, Astrophys. J. Suppl. \textbf{223}, 22
  (2016), \texttt{1511.08519}

\bibitem{Radice:2016dwd}
D.~Radice, F.~Galeazzi, J.~Lippuner, L.F. Roberts, C.D. Ott, L.~Rezzolla, Mon.
  Not. Roy. Astron. Soc. \textbf{460}, 3255 (2016), \texttt{1601.02426}

\bibitem{Bovard:2017mvn}
L.~Bovard, D.~Martin, F.~Guercilena, A.~Arcones, L.~Rezzolla, O.~Korobkin,
  Phys. Rev. \textbf{D96}, 124005 (2017), \texttt{1709.09630}

\bibitem{Perego:2017xth}
A.~Perego, A.~Arcones, D.~Martin, H.~Yasin, JPS Conf. Proc. \textbf{14}, 020810
  (2017)

\bibitem{Foucart:2018gis}
F.~Foucart, M.D. Duez, L.E. Kidder, R.~Nguyen, H.P. Pfeiffer, M.A. Scheel
  (2018), \texttt{1806.02349}

\bibitem{Ardevol-Pulpillo:2018btx}
R.~Ardevol-Pulpillo, H.T. Janka, O.~Just, A.~Bauswein, Mon. Not. Roy. Astron.
  Soc. \textbf{485}, 4754 (2019), \texttt{1808.00006}

\bibitem{Mezzacappa:1993gn}
A.~Mezzacappa, S.W. Bruenn, Astrophys. J. \textbf{405}, 669 (1993)

\bibitem{Roberts.etal:2012}
L.F. {Roberts}, G.~{Shen}, V.~{Cirigliano}, J.A. {Pons}, S.~{Reddy}, S.E.
  {Woosley}, Physical Review Letters \textbf{108}, 061103 (2012),
  \texttt{1112.0335}

\bibitem{Lentz:2012xc}
E.J. Lentz, A.~Mezzacappa, O.E. Bronson~Messer, W.R. Hix, S.W. Bruenn,
  Astrophys. J. \textbf{760}, 94 (2012), \texttt{1206.1086}

\bibitem{Abdikamalov:2014oba}
E.~Abdikamalov, C.~Ott, D.~Radice, L.~Roberts, R.~Haas et~al. (2014),
  \texttt{1409.7078}

\bibitem{Melson:2015}
T.~{Melson}, H.T. {Janka}, R.~{Bollig}, F.~{Hanke}, A.~{Marek},
  B.~{M{\"u}ller}, \apjl \textbf{808}, L42 (2015), \texttt{1504.07631}

\bibitem{Fischer:2016}
T.~{Fischer}, \aap \textbf{593}, A103 (2016), \texttt{1608.05004}

\bibitem{OConnor:2018sti}
E.~O'Connor et~al., J. Phys. \textbf{G45}, 104001 (2018), \texttt{1806.04175}

\bibitem{Pan:2018vkx}
K.C. Pan, C.~Mattes, E.P. O'Connor, S.M. Couch, A.~Perego, A.~Arcones, J. Phys.
  \textbf{G46}, 014001 (2019), \texttt{1806.10030}

\bibitem{Cabezon:2018lpr}
R.M. Cabezón, K.C. Pan, M.~Liebendörfer, T.~Kuroda, K.~Ebinger, O.~Heinimann,
  F.K. Thielemann, A.~Perego, Astron. Astrophys. \textbf{619}, A118 (2018),
  \texttt{1806.09184}

\bibitem{Just.etal:2018}
O.~{Just}, R.~{Bollig}, H.T. {Janka}, M.~{Obergaulinger}, R.~{Glas},
  S.~{Nagataki}, \mnras \textbf{481}, 4786 (2018), \texttt{1805.03953}

\bibitem{Perego:2014qda}
A.~Perego, E.~Gafton, R.~Cabezón, S.~Rosswog, M.~Liebendörfer, Astron.
  Astrophys. \textbf{568}, A11 (2014), \texttt{1403.1297}

\bibitem{Foucart:2015vpa}
F.~Foucart, E.~O'Connor, L.~Roberts, M.D. Duez, R.~Haas, L.E. Kidder, C.D. Ott,
  H.P. Pfeiffer, M.A. Scheel, B.~Szilagyi, Phys. Rev. \textbf{D91}, 124021
  (2015), \texttt{1502.04146}

\bibitem{Shapiro:1983du}
S.L. Shapiro, S.A. Teukolsky, \emph{{Black holes, white dwarfs, and neutron
  stars: The physics of compact objects}} (Wiley, New York, USA, 1983)

\bibitem{Raffelt:2001kv}
G.G. Raffelt, Astrophys. J. \textbf{561}, 890 (2001), \texttt{astro-ph/0105250}

\bibitem{OConnor:Thesis:2012}
E.~{O'Connor}, PhD Thesis  (2012)

\bibitem{Horowitz:2001xf}
C.J. Horowitz, Phys. Rev. \textbf{D65}, 043001 (2002),
  \texttt{astro-ph/0109209}

\bibitem{Tubbs:1975jx}
D.L. Tubbs, D.N. Schramm, Astrophys. J. \textbf{201}, 467 (1975)

\bibitem{Burrows:1981zz}
A.~Burrows, T.J. Mazurek, J.M. Lattimer, Astrophys. J. \textbf{251}, 325 (1981)

\bibitem{Leinson:1997ns}
L.B. Leinson, V.N. Oraevskii, V.B. Semikoz, Sov. J. Nucl. Phys. \textbf{48},
  963 (1988), [Yad. Fiz.48,1513(1988)]

\bibitem{Burrows:2004vq}
A.~Burrows, S.~Reddy, T.A. Thompson, Nucl. Phys. \textbf{A777}, 356 (2006),
  \texttt{astro-ph/0404432}

\bibitem{Hannestad:1997gc}
S.~Hannestad, G.~Raffelt, Astrophys. J. \textbf{507}, 339 (1998),
  \texttt{astro-ph/9711132}

\bibitem{Typel:2009sy}
S.~Typel, G.~Ropke, T.~Klahn, D.~Blaschke, H.H. Wolter, Phys. Rev.
  \textbf{C81}, 015803 (2010), \texttt{0908.2344}

\bibitem{Hempel:2009mc}
M.~Hempel, J.~Schaffner-Bielich, Nucl. Phys. \textbf{A837}, 210 (2010),
  \texttt{0911.4073}

\bibitem{daSilvaSchneider:2017jpg}
A.S. Schneider, L.F. Roberts, C.D. Ott, Phys. Rev. \textbf{C96}, 065802 (2017),
  \texttt{1707.01527}

\bibitem{Gourgoulhon:2000nn}
E.~Gourgoulhon, P.~Grandclement, K.~Taniguchi, J.A. Marck, S.~Bonazzola,
  Phys.Rev. \textbf{D63}, 064029 (2001), \texttt{gr-qc/0007028}

\bibitem{Radice:2012cu}
D.~Radice, L.~Rezzolla, Astron. Astrophys. \textbf{547}, A26 (2012),
  \texttt{1206.6502}

\bibitem{Radice:2013hxh}
D.~Radice, L.~Rezzolla, F.~Galeazzi, Mon.Not.Roy.Astron.Soc. \textbf{437}, L46
  (2014), \texttt{1306.6052}

\bibitem{Radice:2013xpa}
D.~Radice, L.~Rezzolla, F.~Galeazzi, Class.Quant.Grav. \textbf{31}, 075012
  (2014), \texttt{1312.5004}

\bibitem{Radice:2015nva}
D.~Radice, L.~Rezzolla, F.~Galeazzi, ASP Conf. Ser. \textbf{498}, 121 (2015),
  \texttt{1502.00551}

\bibitem{Radice:2018xqa}
D.~Radice, A.~Perego, S.~Bernuzzi, B.~Zhang, Mon. Not. Roy. Astron. Soc.
  \textbf{481}, 3670 (2018), \texttt{1803.10865}

\bibitem{Bacca:2011qd}
S.~Bacca, K.~Hally, M.~Liebendorfer, A.~Perego, C.J. Pethick, A.~Schwenk,
  Astrophys. J. \textbf{758}, 34 (2012), \texttt{1112.5185}

\bibitem{Rrapaj:2014yba}
E.~Rrapaj, J.W. Holt, A.~Bartl, S.~Reddy, A.~Schwenk, Phys. Rev. \textbf{C91},
  035806 (2015), \texttt{1408.3368}

\bibitem{Roberts:2016mwj}
L.F. Roberts, S.~Reddy, Phys. Rev. \textbf{C95}, 045807 (2017),
  \texttt{1612.02764}

\bibitem{Guo:2019cvs}
G.~Guo, G.~Martínez-Pinedo (2019), \texttt{1905.13634}

\end{thebibliography}

\end{document}